\documentclass[]{aa}  

\usepackage{graphicx}
\usepackage{txfonts}
\usepackage[]{hyperref}
\usepackage{comment}
\usepackage{natbib}
\usepackage{xcolor}
\usepackage{overpic}
\bibpunct{(}{)}{;}{a}{}{,}

\begin{document} 

   \title{A z=1.85 galaxy group in CEERS: Evolved, dustless, massive intra-halo light and a brightest group galaxy in the making}

    
      \date{Received February 20, 2023; accepted May 15, 2023}

  \abstract{We present CEERS JWST/NIRCam imaging of a massive galaxy group at z=1.85, to explore the early JWST view on massive group formation in the distant Universe. The group contains $\gtrsim$16 members (including six spectroscopic confirmations) down to log$_{10}$(M$_{\star}$/M$_{\odot}$)=8.5, including the brightest group galaxy (BGG) in the process of actively assembling at this redshift. The BGG is comprised of multiple merging components extending $\sim$3.6" (30~kpc) across the sky. The BGG contributes 69\% of the group's total galactic stellar mass, with one of the merging components containing 76\% of the total mass of the BGG and a star formation~rate~$>$1810M$_{\odot}$/yr. Most importantly, we detected intra-halo light (IHL) in several HST and JWST/NIRCam bands, allowing us to construct a state-of-the-art rest-frame UV-NIR spectral energy distribution of the IHL for the first time at this high redshift. This allows stellar population characterisation of both the IHL and member galaxies, as well as the morphology distribution of group galaxies versus their star formation activity when coupled with Herschel data. We created a stacked image of the IHL, giving us a sensitivity to extended emission of 28.5 mag/arcsec$^{2}$ at rest-frame 1~$\mu$m. We find that the IHL is extremely dust poor (A$_{v}$$\sim$0), containing an evolved stellar population of log$_{10}$(t$_{50}$/yr)=8.8, corresponding to a formation epoch for 50\% of the stellar material 0.63~Gyr before z=1.85. There is no evidence of ongoing star formation in the IHL. The IHL in this group at z=1.85 contributes $\sim$10\% of the total stellar mass, comparable with what is observed in local clusters. This suggests that the evolution of the IHL fraction is more self-similar with redshift than predicted by some models, challenging our understanding of IHL formation during the assembly of high-redshift clusters. JWST is unveiling a new side of group formation at this redshift, which will evolve into Virgo-like structures in the local Universe.} 


   \author{
   	Rosemary T. Coogan \inst{1}\thanks{rosemary.coogan@cea.fr}
   	\and
          Emanuele Daddi \inst{1}
          \and
         Aur{\'e}lien Le Bail \inst{1}
          \and
         David Elbaz\inst{1}
         \and
         Mark Dickinson \inst{2}
         \and
	Mauro Giavalisco \inst{3}
	 \and
         Carlos G\'omez-Guijarro \inst{1}
	 \and
         Alexander de la Vega \inst{4}
         \and
         Micaela Bagley \inst{5}
         \and 
         Steven L. Finkelstein \inst{5}
         \and
         Maximilien Franco \inst{5}
         \and
         Asantha R. Cooray \inst{6}
         \and
         Peter Behroozi \inst{7, 8}
         \and
         Laura Bisigello \inst{9, 10}
         \and
         Caitlin M. Casey \inst{5}
         \and
         Laure Ciesla \inst{11}
         \and
         Paola Dimauro  \inst{12}
         \and
	Alexis Finoguenov \inst{13}
         \and
         Anton M. Koekemoer \inst{14}              
	\and
         Ray A. Lucas \inst{14}
         \and
        Pablo G. P\'erez-Gonz\'alez \inst{15}
         \and
         {L. Y. Aaron} {Yung} \inst{16}\thanks{NASA Postdoctoral Fellow}
         \and 
         Pablo Arrabal Haro \inst{2}
         \and
         Jeyhan S. Kartaltepe \inst{17}
          \and
         Shardha Jogee \inst{5}
         \and
         Casey Papovich \inst{18}
         \and
         Nor Pirzkal \inst{19}
         \and 
         Stephen M.~Wilkins \inst{20, 21}
         }

   \institute{Universit{\'e} Paris-Saclay, Universit{\'e} Paris Cit{\'e}, CEA, CNRS, AIM, 91191, Gif-sur-Yvette, France
             \and
             NSF's National Optical-Infrared Astronomy Research Laboratory, 950 N. Cherry Ave., Tucson, AZ 85719, USA
	    \and
             University of Massachusetts Amherst, 710 North Pleasant Street, Amherst, MA 01003-9305, USA
             \and
             Department of Physics and Astronomy, University of California, 900 University Ave, Riverside, CA 92521, USA
             \and
             Department of Astronomy, The University of Texas at Austin, Austin, TX, USA
	     \and
             Department of Physics \& Astronomy, University of California, Irvine, 4129 Reines Hall, Irvine, CA 92697, USA
             \and
             Department of Astronomy and Steward Observatory, University of Arizona, Tucson, AZ 85721, USA
             \and
             Division of Science, National Astronomical Observatory of Japan, 2-21-1 Osawa, Mitaka, Tokyo 181-8588, Japan
             \and
             Dipartimento di Fisica e Astronomia "G.Galilei", Universit\'a di Padova, Via Marzolo 8, I-35131 Padova, Italy
             \and
             INAF--Osservatorio Astronomico di Padova, Vicolo dell'Osservatorio 5, I-35122, Padova, Italy
             \and
             Aix Marseille Univ, CNRS, CNES, LAM Marseille, France
             \and
             INAF - Osservatorio Astronomico di Roma, via di Frascati 33, 00078, Monte Porzio Catone, Italy
             \and
             Department of Physics, University of Helsinki, Gustaf H\"allstr\"omin katu 2, 00014 Helsinki, Finland
             \and
             Space Telescope Science Institute, 3700 San Martin Dr., Baltimore, MD 21218, USA
             \and
             Centro de Astrobiolog\'{\i}a (CAB), CSIC-INTA, Ctra. de Ajalvir km 4, Torrej\'on de Ardoz, E-28850, Madrid, Spain
             \and
             Astrophysics Science Division, NASA Goddard Space Flight Center, 8800 Greenbelt Rd, Greenbelt, MD 20771, USA
             \and
             Laboratory for Multiwavelength Astrophysics, School of Physics and Astronomy, Rochester Institute of Technology, 84 Lomb Memorial Drive, Rochester, NY 14623, USA
             \and
             Department of Physics and Astronomy, Texas A\&M University, College Station, TX, 77843-4242 USA
             \and
             ESA/AURA Space Telescope Science Institute
             \and
             Astronomy Centre, University of Sussex, Falmer, Brighton BN1 9QH, UK
             \and
             Institute of Space Sciences and Astronomy, University of Malta, Msida MSD 2080, Malta
             }

   \titlerunning{IHL in a z=1.85 galaxy group in CEERS}
   \maketitle
%

\section{Introduction}
Clusters and groups of galaxies open up a wealth of science. These impressive structures collapse from high density fluctuations in the early Universe and, as such, the number density of clusters and their spatial distribution are highly sensitive to the underlying cosmology (e.g. \citealt{ref:S.Allen2011}), and thus dark matter and dark energy density parameters. In addition, galaxy (proto-)clusters and groups are the unambiguous formation sites of the first massive galaxies, and can constrain the elusive processes leading to the formation of massive red, `dead' galaxies. The cluster environment has a clear impact on galaxy properties in the local Universe, with cluster galaxies depleting the molecular gas that fuels star formation and becoming quenched, early-type galaxies far more rapidly than equivalent field galaxies (e.g. \citealt{ref:A.Dressler1980}, \citealt{ref:G.Kauffmann2004}). However, the physical processes driving this accelerated evolution are not well understood, and are observable only by detecting galaxy clusters at high redshift (z$\gtrsim$1).

In addition to the galaxies themselves, one of the most revealing and unique observational features of galaxy groups and clusters is the diffuse intra-halo light (IHL). The IHL is continuum emission from stellar material that fills the space between galaxies in these dense environments \citep{ref:F.Zwicky1957}, and it does not appear to be directly associated with individual galaxies. The presence of IHL is one of the most revealing signatures of the hierarchical assembly of clusters through the (environment-dependent) accretion and interaction of galaxies (see e.g. \citealt{ref:C.Mihos2016, ref:E.Contino2021} for reviews), and this diffuse light is composed of a substantial fraction of stars, between 5–20\% of the total stars in local clusters (e.g. \citealt{ref:J.Krick2007, ref:C.Burke2015, ref:M.Montes2018, ref:Y.Jimenez2018}, see \citealt{ref:C.MartinezLombilla2023} for IHL in a local galaxy group). 

However, the stage of cluster assembly at which the IHL forms, and the processes that give rise to this diffuse light are highly uncertain. One possible source of IHL would be stellar material stripped from galaxies during merger events. This can occur during galaxy-galaxy interactions in galaxy groups, leading to tidal streams which are then mixed with the IHL when the group itself is accreted (e.g. \citealt{ref:C.Rudick2006, ref:C.Rudick2009}), or from accreted galaxies directly interacting with cluster members or the cluster potential itself (\citealt{ref:C.Conroy2007, ref:C.Purcell2007, ref:E.Contini2014}). Furthermore, galaxy interactions in the cluster core may lead to the ejection of stellar material into the diffuse surroundings, in the process of assembling the brightest cluster (or group) galaxy (BGG e.g. \citealt{ref:G.Murante2007}). Additionally, in situ star formation may occur in the intra-cluster medium and contribute towards the IHL, if cold gas is also stripped from the galaxies, along with the stellar material (\citealt{ref:E.Puchwein2010, ref:T.Webb2015, ref:M.Mcdonald2016}).

Observationally, the IHL is ubiquitous in low-redshift galaxy clusters, and has been well characterised in the redshift range 0$<$z$<$0.5, where the evolution of IHL properties (e.g. colour) is found to be relatively small (e.g. \citealt{ref:E.Giallongo2014}). The IHL is observed to have colours that suggest the stellar material is relatively evolved and therefore formed at early epochs, similar to the stellar properties of quiescent, red-sequence galaxies within clusters. The IHL has also been shown to be an excellent tool to trace the smooth profile of the dark matter halos and mass in clusters (\citealt{ref:M.Montes2018, ref:G.Mahler2022}), suggesting that the IHL itself is also virialised, having formed over an extended period of time as galaxies and halos are continuously accreted.

On the other hand, the picture is much less clear at intermediate-high redshifts, making a prediction as to the origin of the IHL in its early stages of creation extremely challenging. There have however been several attempts at modelling the complex accretion history of clusters in an attempt to do just this. \citet{ref:C.Rudick2011} use collisionless simulations to model the various processes described above, and predict that the stellar mass contained in the IHL will decrease rapidly with increasing redshift, become significantly less relevant beyond z=1 compared with the local Universe, and become entirely negligible at z$>$2. Furthermore, \citet{ref:E.Contini2014} come to a similar conclusion that the IHL forms at z$<$1, using a semi-analytical model of galaxy formation, coupled with merger trees extracted from N-body simulations of groups and clusters. \citet{ref:E.Contini2018} model the ratio between IHL and brightest cluster galaxy mass as a function of redshift, and predict that the IHL-to-brightest cluster galaxy mass ratio will be around 2-4\% at z=1-2, if both merging galaxies and stellar stripping contribute to the IHL. Furthermore, the UniverseMachine of \citet{ref:P.Behroozi2019} predict that the fraction of IHL mass to total stellar mass is invariant with redshift between z~=~0 and z~=~2, with the IHL contributing $\sim$10\% to the total stellar mass for a halo mass of log$_{10}$(M$_{\rm halo, peak}/M_{\odot}$) = 13.4. We have very few observational constraints on these processes at z$>$1, especially at longer wavelengths than what can be seen with the \textit{Hubble} Space Telescope (HST). Recently however, \citet{ref:H.Joo2023} studied the IHL in ten 1~$\leq$~z~$\leq$~1.8 clusters based on deep infrared continuum data, and found a mean IHL fraction of 17\%. Detection of the IHL at these higher redshifts, and furthermore accurate characterisation of stellar properties (such as age), are crucial in order for us to understand both the formation epoch and mechanisms of the IHL.

\begin{figure}
\includegraphics[width=0.48\textwidth, clip, trim=0cm 0cm 0cm 0.cm]{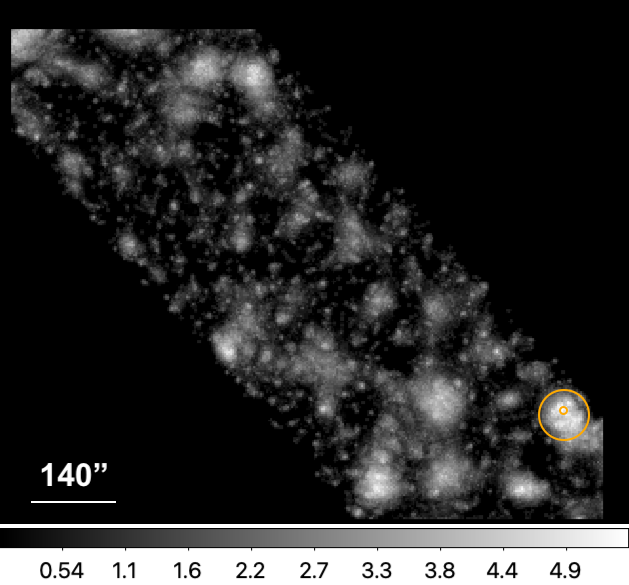}
\caption{Overdensity map of the EGS field, in units of sigma. The position of the group at z=1.85 is indicated by the small orange circle (r$\sim$5") near the lower right, and the larger orange circle is only there to guide the eye to the high density part of the map.}
\label{fig:overdensity}
\end{figure}

\begin{figure*}
\begin{minipage}{\textwidth}
\includegraphics[width=0.48\textwidth, clip, trim=2cm 6cm 1.5cm 0.cm]{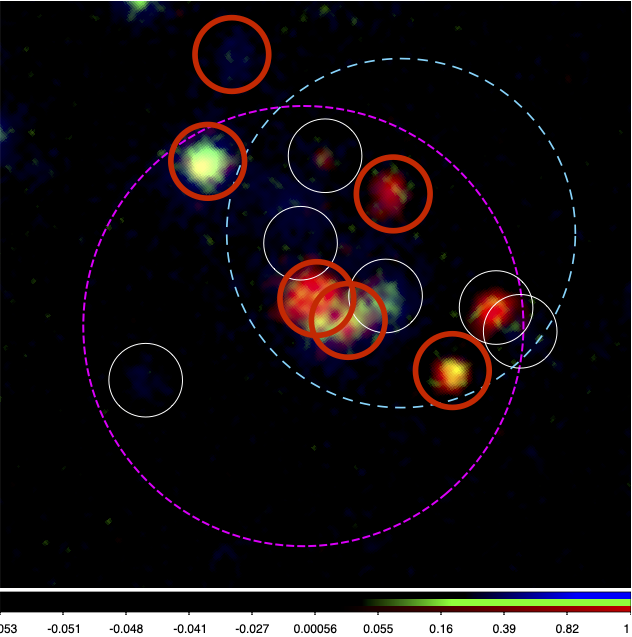}\hfill
\hspace{2.9mm}\includegraphics[width=0.48\textwidth, clip, trim=0.4cm 1.8cm 1cm 0.cm]{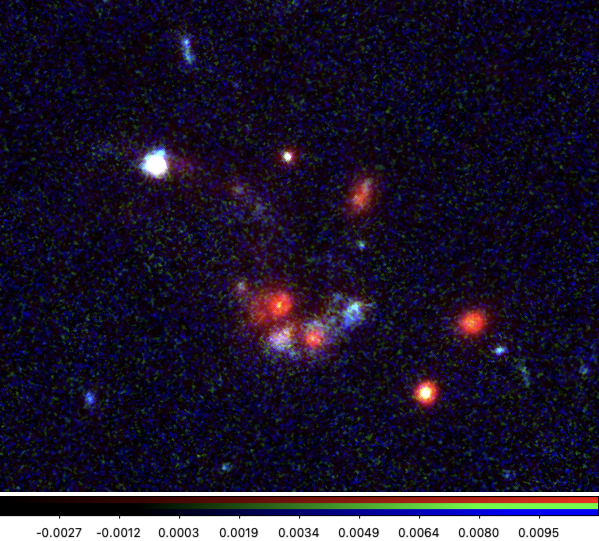}\hfill
\end{minipage}
\begin{minipage}{\textwidth}
\includegraphics[width=0.48\textwidth, clip, trim=0.5cm 0cm 0.6cm 0cm]{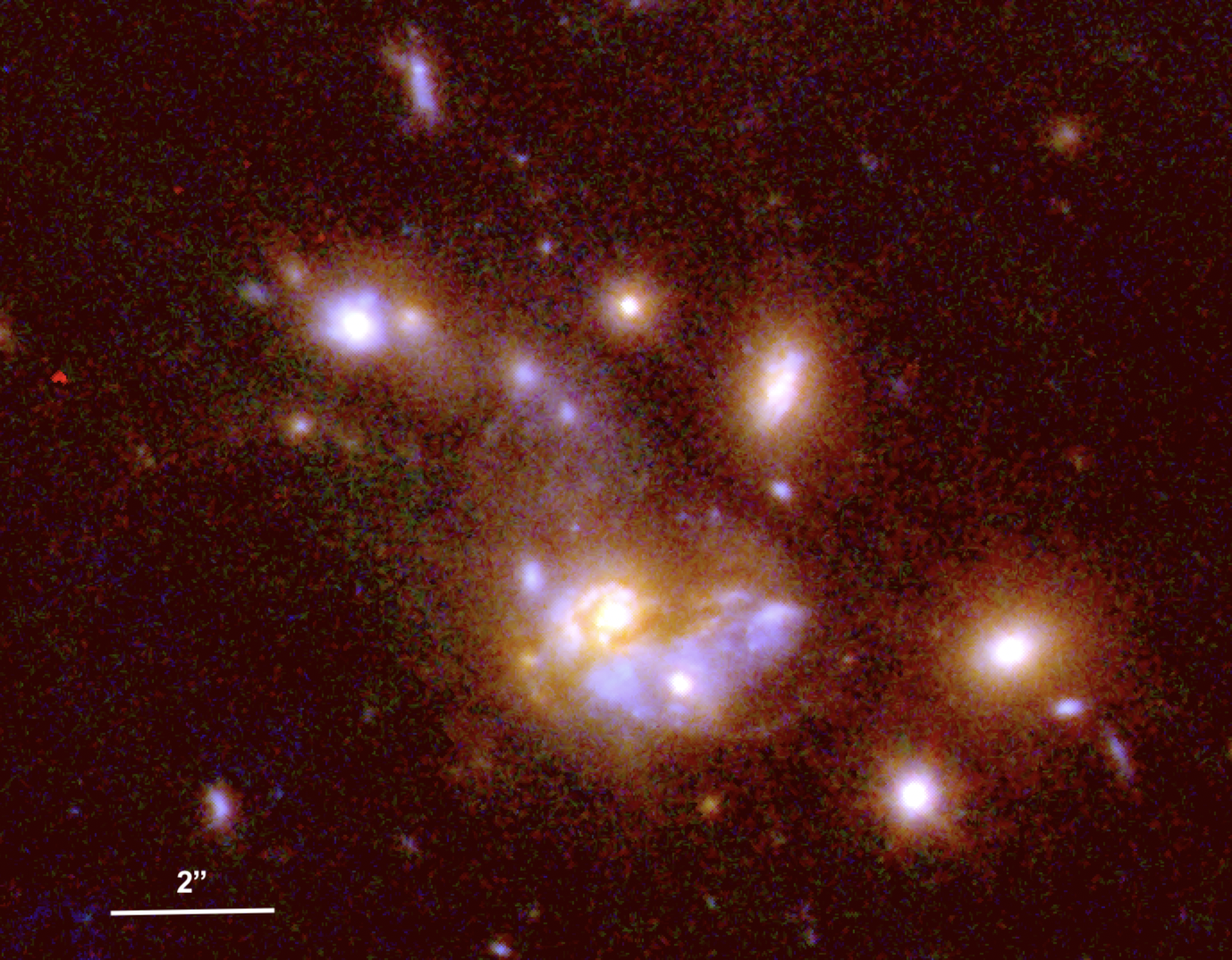}\hfill
\includegraphics[width=0.48\textwidth, clip, trim=0.cm 1.5cm 4cm 0.7cm]{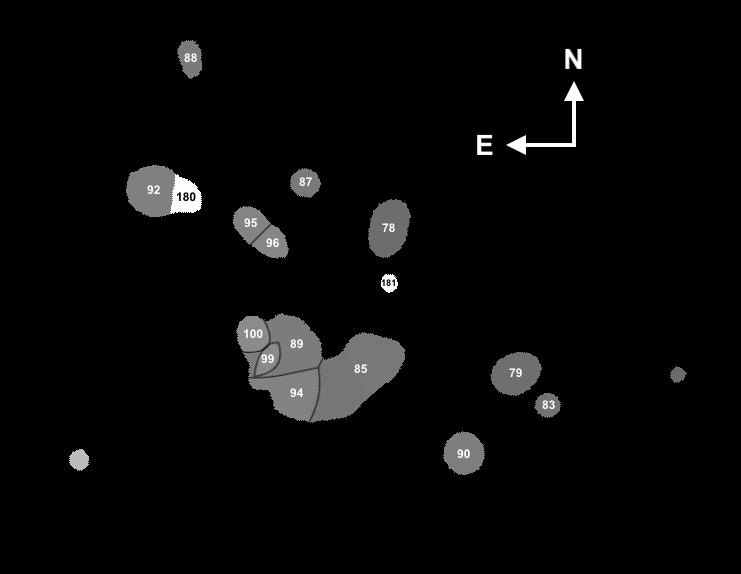}\hfill
\end{minipage}
\caption{Different views of the galaxy group at z=1.85. Upper left: Subaru GZKs colour image of the z=1.85 group in the EGS field. Members of the \citet{ref:M.Stefanon2017} catalogue between 1.5~$<$~z$_{\rm phot}$~$<$~2.0 are shown by small white and red circles. The red circles indicate that spectroscopic HST grism confirmations are available (see Table~\ref{tab:zs}), white is photometric only. The r=5.98" aperture that defines the 5.3$\sigma$ overdensity is shown in magenta (south), but we also show the more concentrated r=4.75" aperture that gives a significance of 5.2$\sigma$ in cyan (north). Upper right: HST F606W-F814W-F160W image of the group. Lower left: NIRCam F115W-F200W-F444W colour image of the group. Lower right: galaxy segmentation maps identified by SExtractor, used to measure galaxy photometry in all HST and JWST bands (borders have been drawn between some segmentation maps for clarity). We emphasise that the segmentation maps shown here are those derived from SExtractor. The radially grown mask used to exclude galaxies when making the IHL measurement is larger than the segmentation maps shown here (see Figs~\ref{fig:cutouts}, \ref{fig:stacked}). The lower-left galaxy is not considered as it falls outside of the cyan aperture, as discussed in the text.}
\label{fig:bcg}
\end{figure*}

In this paper, we use JWST Near InfraRed Camera (NIRCam) observations, along with complementary data from HST, for example, to investigate the physical properties of the IHL and member galaxies in a galaxy group at z=1.85. In Sections \ref{sec:obs} and \ref{sec:methods}, we outline the datasets being used for this analysis, and describe our methodology for identifying the group and performing spectral energy distribution (SED) fitting on both the member galaxies and the IHL. In Section \ref{sec:results}, we present our results and discuss the implications for understanding the properties and formation mechanisms of the IHL at high redshift. We summarise in Section~\ref{sec:conc}. Throughout the paper, we use a $\Lambda$CDM cosmology with H$_{0}$=70kms$^{-1}$Mpc$^{-1}$, $\Omega_{M}$=0.3 and $\Omega_{\Lambda}$=0.7, and a \citet{ref:G.Chabrier2003} initial mass function.


\section{Observations}
\label{sec:obs}

\subsection{JWST/CEERS data}
The Cosmic Evolution Early Release Science (CEERS\footnote{\url{https://ceers.github.io/}}) survey is one of the 13 programmes chosen to be carried out near the beginning of JWST's \citep{ref:J.Gardner2006} scientific operation, where the observations are made available to the public immediately, in order to demonstrate both data reduction techniques and achievable new scientific goals. CEERS will consist of 10 NIRCam pointings over $\sim$100~arcmin$^{2}$ in the Extended Groth Strip (EGS), with the Near InfraRed Spectrograph instrument (NIRSpec) covering six of these pointings, and Mid InfraRed Instrument (MIRI) coverage for a subset of eight pointings. Here, we use a single pointing, CEERS3, of the first JWST/NIRCam observations taken as part of CEERS on the 21 June 2022. Data were obtained in the short wavelength channel F115W, F150W, and F200W filters, and long-wavelength channel F277W, F356W, F410M, and F444W filters. The total exposure time for pixels observed was typically 2835s per filter, although the exposure time was double for the F115W filter. An initial reduction of the NIRCam images was made using version 1.5.3 of the JWST Calibration Pipeline1, with some custom modifications. The reduction steps are described in more detail in \citet{ref:S.Finkelstein2022} and \citet{ref:M.Bagley2022}, with full details to be given in the survey overview paper (Finkelstein et al. in prep.). All images are at a pixel scale of 0.03"/pix. We use the publicly released v0.5 of the CEERS3 data, without the local background removal, which could inadvertently over-subtract the IHL where present.

\subsection{Hubble Space Telescope}
We use the publicly available HST data products version 1.9, available through CEERS\footnote{\url{https://ceers.github.io/releases.html\#hdr1}}, for the ACS/WFC F606W, F814W, and WFC3/IR F125W filters. These mosaics are derived from HST archival data, but with improved calibration compared to the default pipeline products, and have astrometry tied to Gaia-ERD3. As described in the accompanying data release, the mosaics are created from the combination of HST programmes 10134, 12063, 12099, 12167, 12177, 12547, 13063, and 13792, and the reduction and calibration follow a similar procedure to those described in \citet{ref:A.Koekemoer2011}. Mosaics are at a pixel scale of 0.03"/pix.

\begin{figure}
\centering
\includegraphics[width=0.5\textwidth, clip, trim=0.7cm 0.7cm 0cm 0cm]{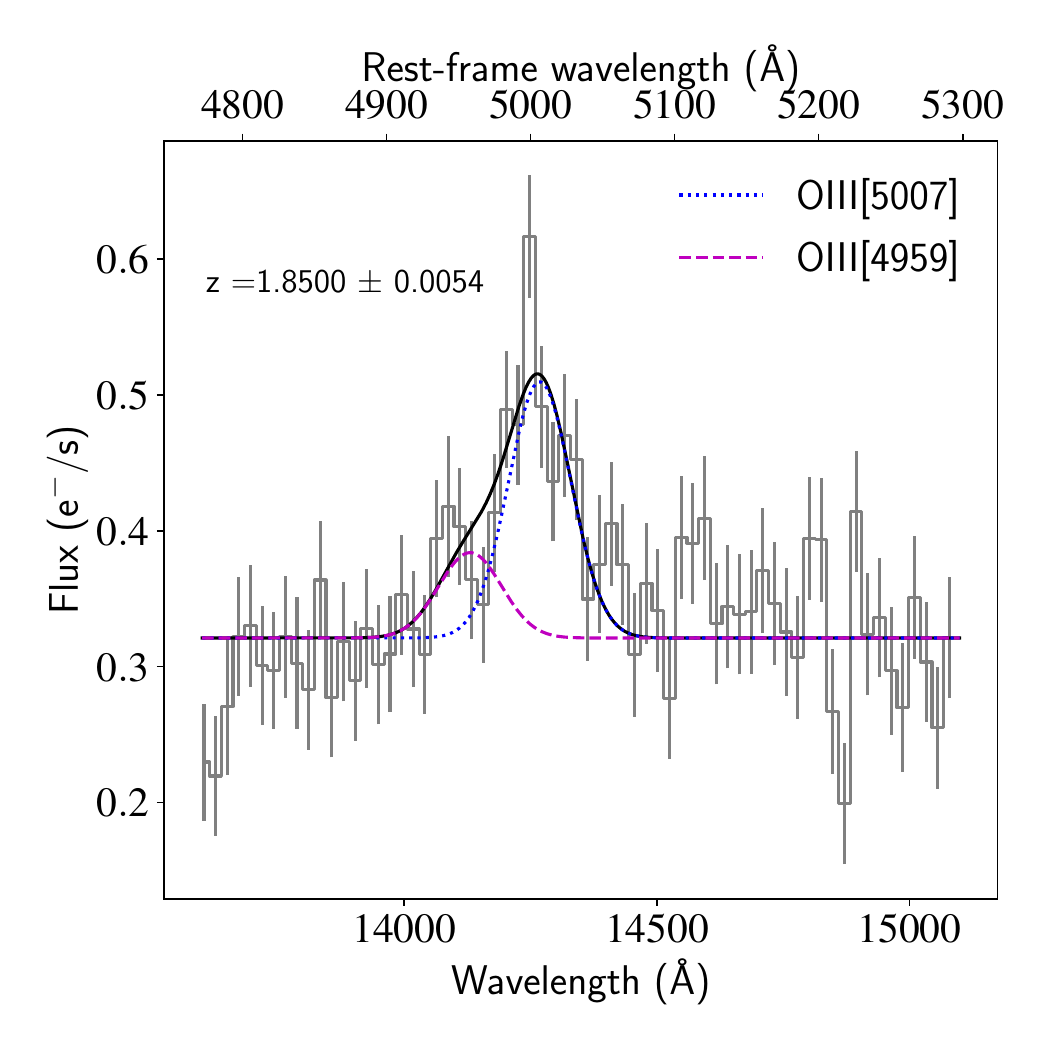}
\caption{1D HST G141W grism spectrum for galaxy seg. number 89. The [OIII]5007 and [OIII]4959 Gaussian fits are shown by the coloured dotted lines, as indicated in the legend, with the sum of the two shown by the black solid line.}
\label{fig:grism_z}
\end{figure}

\section{Methods}
\label{sec:methods} 

\subsection{Searching for overdensities in the EGS field}
In order to systematically search for overdensities of galaxies that may indicate galaxy groups or clusters in the EGS field, we use the \citet{ref:M.Stefanon2017} catalogue. A. Le Bail et al. in prep. have created a compilation of the positions, photometric redshifts and photometric data from \citet{ref:M.Stefanon2017}, \citet{ref:J.Geach2017} and \citet{ref:J.Zavala2017}, and combined this with super-deblended far-infrared (FIR) fluxes, which we subsequently refer to as the `FIR catalogue', (with associated CANDELS IDs from \citealt{ref:M.Stefanon2017}). We searched for overdensities of galaxies within the photometric redshift bins [1.2, 1.6], [1.5, 2.0], [1.9, 2.5], [2.4, 3.1], and [2.4, 3.8], for galaxies with stellar mass M$_{\star}\geq$10$^{7}$M$_{\odot}$, in the \citet{ref:M.Stefanon2017} catalogue. We systematically searched around points in the EGS field corresponding to a regular grid with 3" spacing, and measured the number of galaxies present in the catalogue in the given redshift bin, within a certain distance of that centre position. At each position, we considered galaxies within 12 circular apertures of radii logarithmically spaced between 3" and 38". We then measured the expected number of background galaxies within a circular aperture of the same size, given the stellar mass and redshift range, based on the entire EGS field present in the catalogue. The significance of the overdensity of galaxies found at each position was calculated using Poissonian statistics, based on the relative number of galaxies compared to background expectations. We will present this methodology and subsequent results for larger fields in an upcoming paper.

Using this technique, we discover a 5.3$\sigma$ overdensity of galaxies around RA~=~14:19:00.21, Dec~=~+52:49:47.80 (Fig.~\ref{fig:overdensity}), containing 11 galaxies in the catalogue within an aperture of radius r=5.98". This aperture is shown in Fig.~\ref{fig:bcg} (upper left) in magenta. We note that we also identify the overdensity at 5.2$\sigma$ in a smaller aperture, r=4.8", containing nine galaxies, as shown by the cyan circle in Fig.~\ref{fig:bcg} (upper left).

\begin{figure*}
\centering
\begin{minipage}{\textwidth}
\includegraphics[width=0.24\textwidth, clip, trim=1cm 0.7cm 1cm 0.7cm]{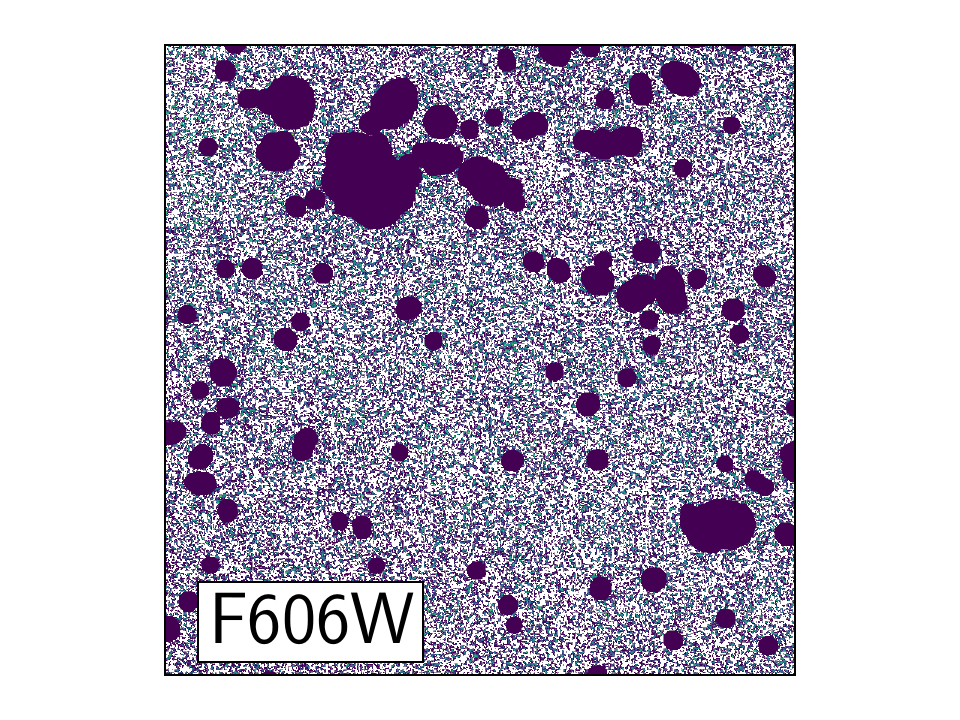}\hfill
\includegraphics[width=0.24\textwidth, clip, trim=1cm 0.7cm 1cm 0.7cm]{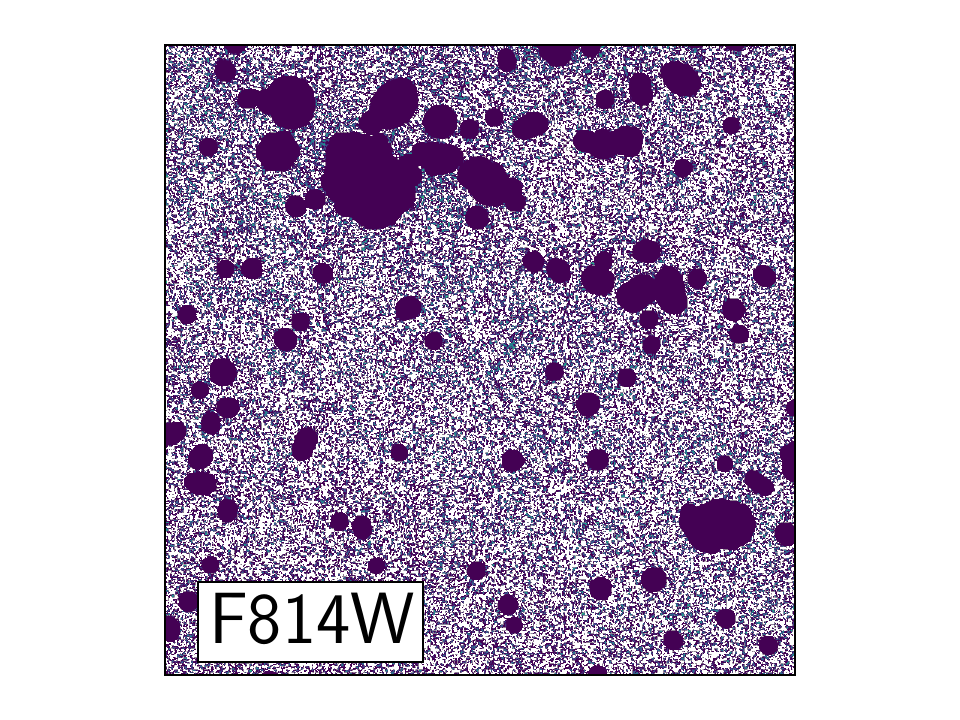}\hfill
\includegraphics[width=0.24\textwidth, clip, trim=1cm 0.7cm 1cm 0.7cm]{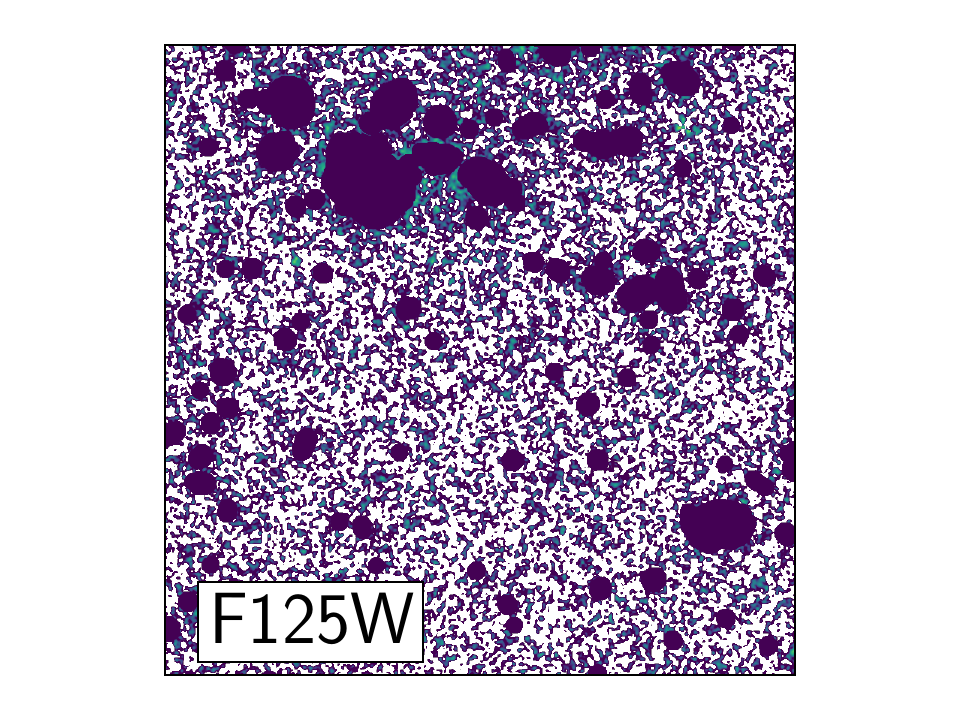}\hfill
\includegraphics[width=0.24\textwidth, clip, trim=1cm 0.7cm 1cm 0.7cm]{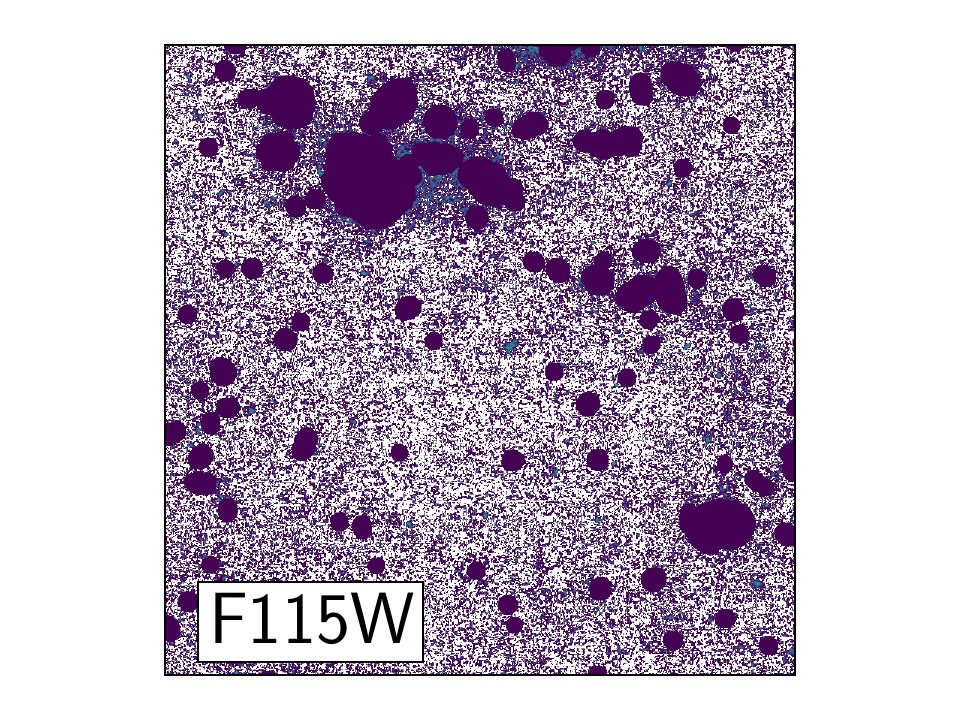}\hfill
\end{minipage}
\begin{minipage}{\textwidth}
\includegraphics[width=0.24\textwidth, clip, trim=1cm 0.7cm 1cm 0.7cm]{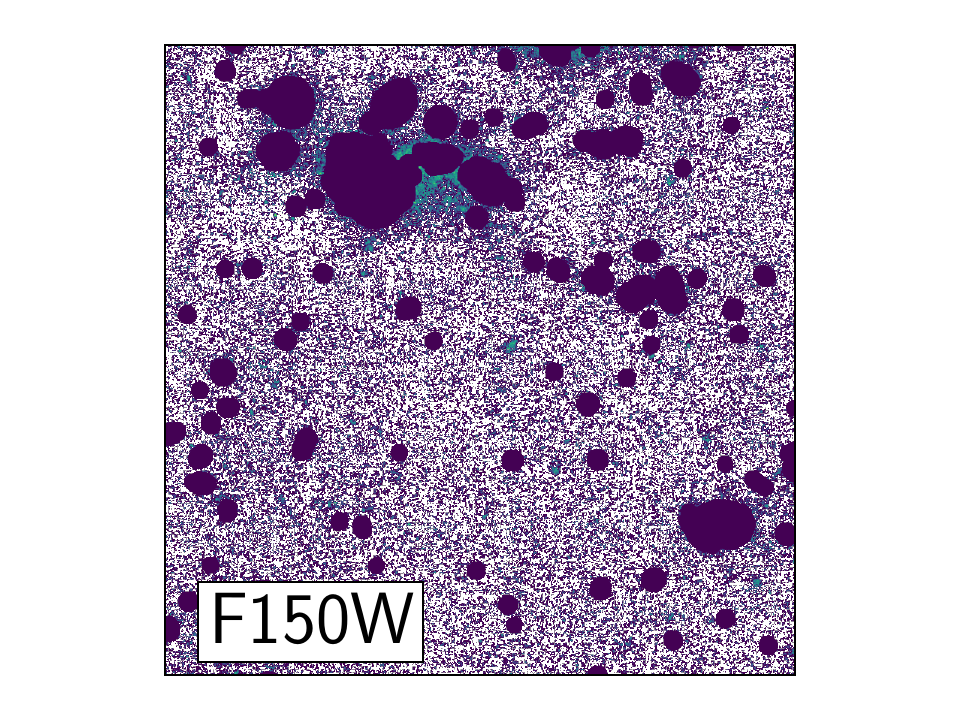}\hfill
\includegraphics[width=0.24\textwidth, clip, trim=1cm 0.7cm 1cm 0.7cm]{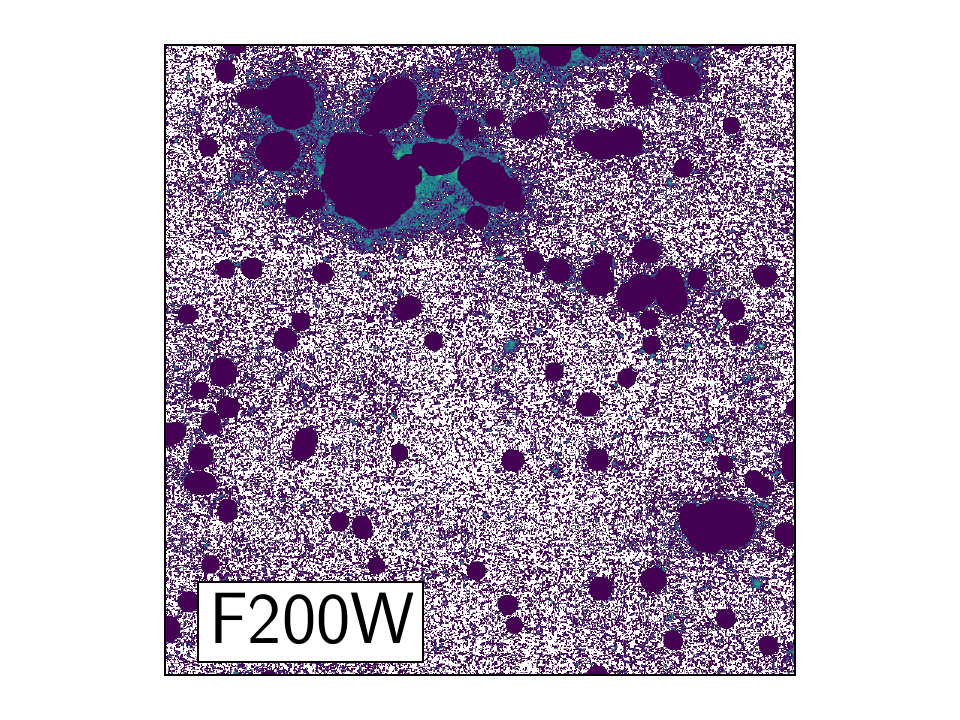}\hfill
\includegraphics[width=0.24\textwidth, clip, trim=1cm 0.7cm 1cm 0.7cm]{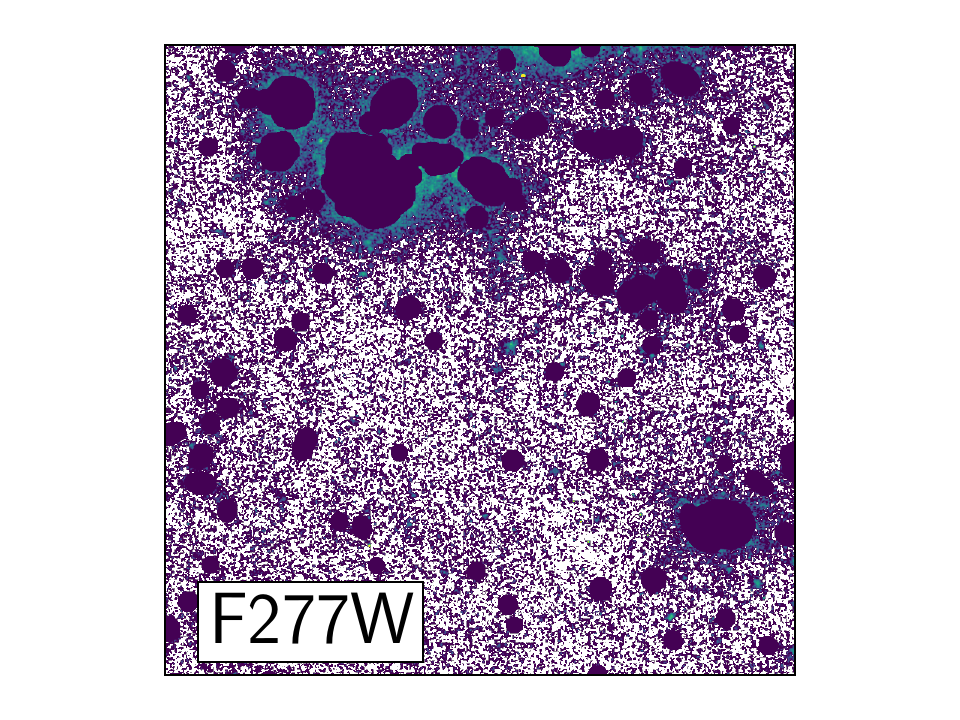}\hfill
\includegraphics[width=0.24\textwidth, clip, trim=1cm 0.7cm 1cm 0.7cm]{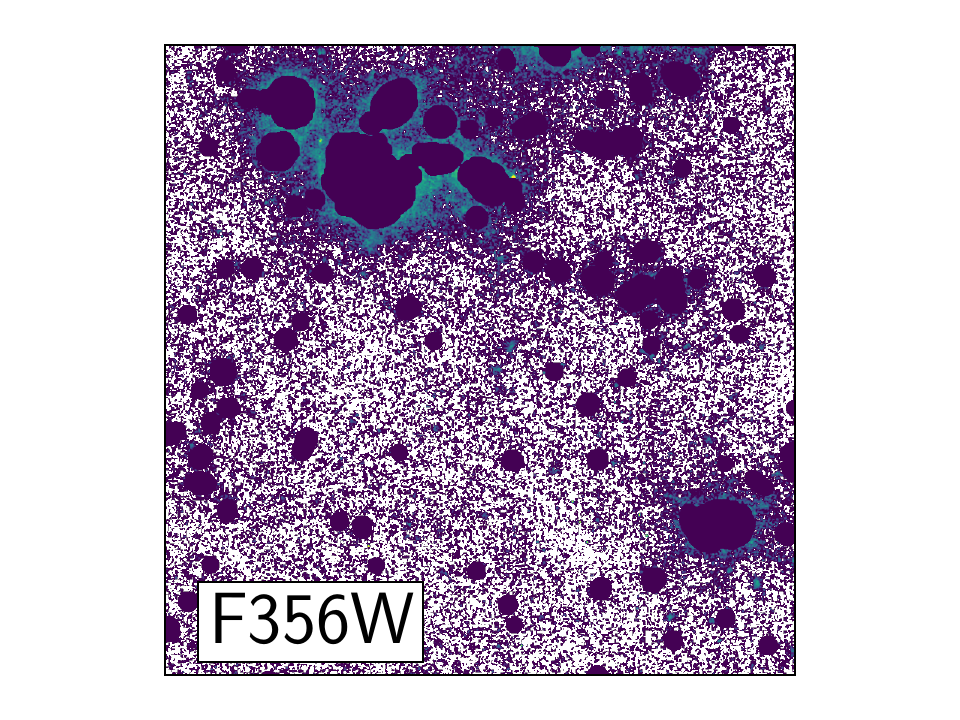}\hfill
\end{minipage}
\begin{minipage}{\textwidth}
\includegraphics[width=0.24\textwidth, clip, trim=1cm 0.7cm 1cm 0.7cm]{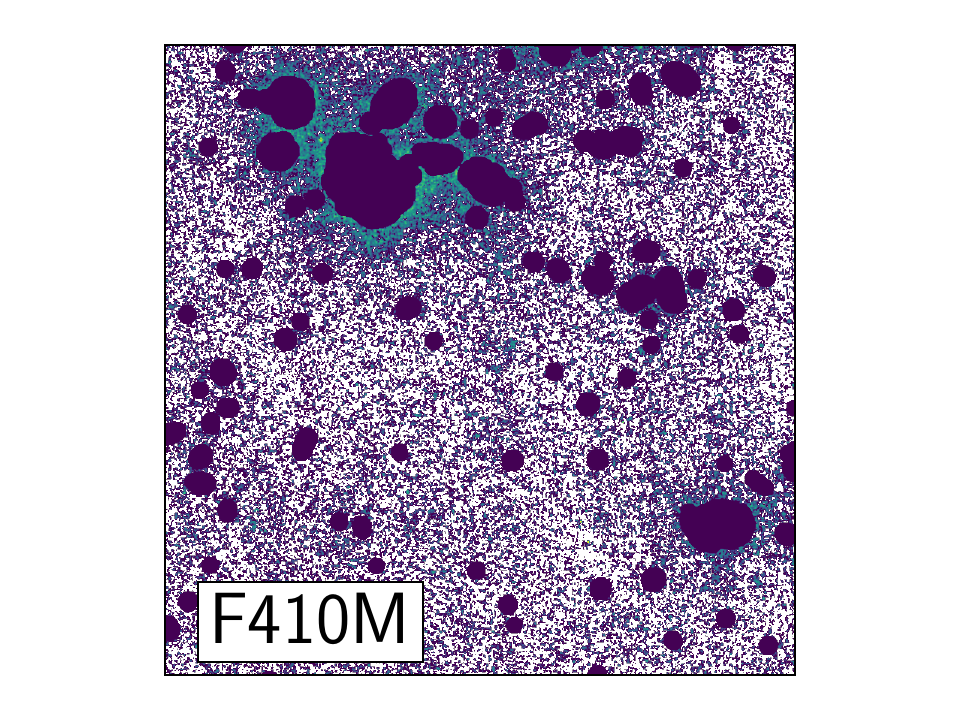}\hfill
\includegraphics[width=0.24\textwidth, clip, trim=1cm 0.7cm 1cm 0.7cm]{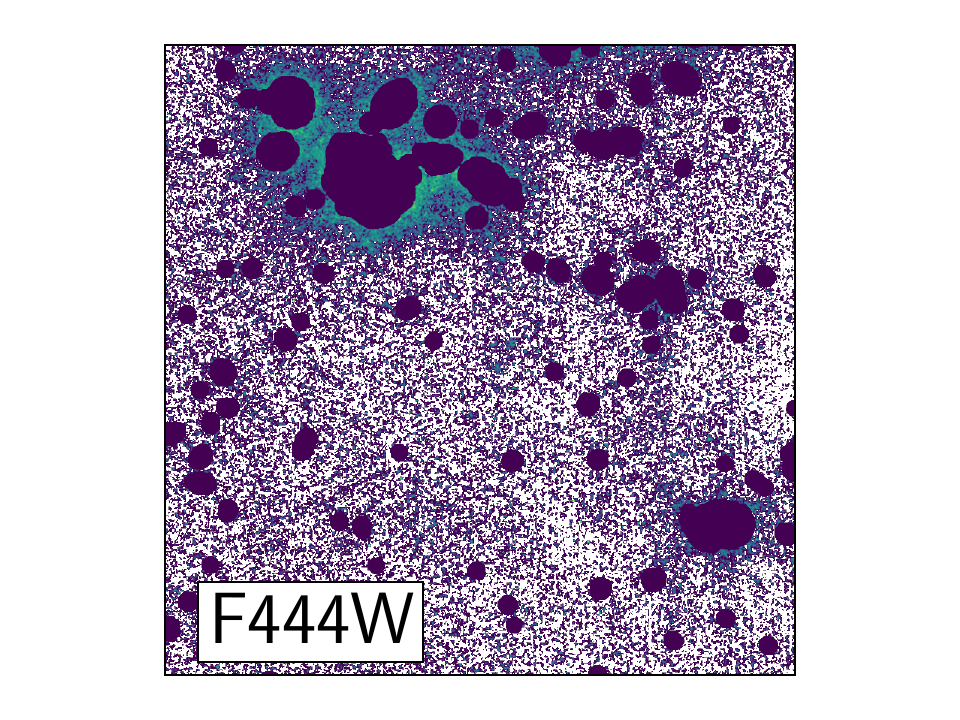}\hfill
\raisebox{0.0cm}{\includegraphics[width=0.24\textwidth, clip, trim=1cm 0.7cm 1cm 0.7cm]{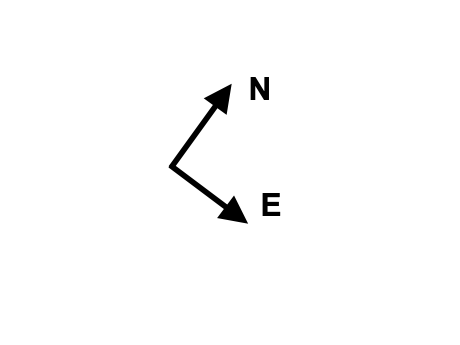}}\hfill
\raisebox{0.0cm}{\includegraphics[width=0.24\textwidth, clip, trim=1cm 0.7cm 1cm 0.7cm]{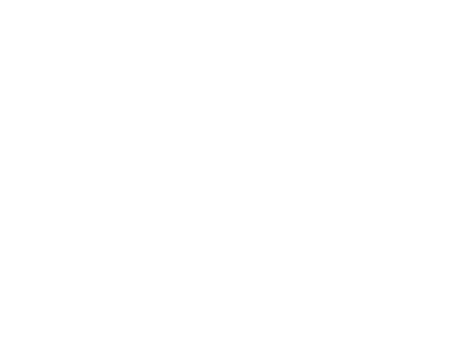}}\hfill
\end{minipage}
\caption{26" x 26" cutouts of the group and the surrounding region, with the master mask overlaid (it is important to note the different orientation). This mask was used for photometric measurements of the IHL. From top to bottom, left to right, the three HST bands are shown, followed by the seven JWST bands, with the filters indicated in the bottom left. We show a large region of the sky to demonstrate the masking of galaxies and lack of diffuse light outside of the group. To the north of the group is a large (lower-redshift) galaxy with a tidal stream that we do not believe is part of the group. We therefore do not show this part of the sky, for clarity.}
\label{fig:cutouts}
\end{figure*}

\subsection{Spectroscopic redshifts}

In order to spectroscopically confirm the redshift of this structure, we used the publicly available 3DHST G141W grism spectroscopy over this region of the sky \citep{ref:R.Skelton2014, ref:I.Momcheva2016}. We were able to confirm a redshift of z$_{\rm spec}$=1.85 for six galaxies in the overdensity (marked in red in Fig.~\ref{fig:bcg} upper left) - including two galaxies which appear to be closely physically associated (segmentation maps 89 and 94, see Table~\ref{tab:zs} and Section~\ref{disc:BCG}). We note that the slightly higher apparent redshift for galaxy 94 most likely arises from the small offset between the grism placement and the bright core seen in Fig~\ref{fig:bcg}. Redshift derivation was based on the fitting of [OIII]5007 and [OIII]4959 lines, keeping in mind the 3DHST collaboration's spectroscopic redshift estimations and the photometric redshift measurements. All lines were fit with a Gaussian profile, with the ratio of the OIII[5007] to OIII[4959] amplitudes equal to 3, and the constraint of a common full width half maximum (FWHM). Errors were derived from Monte Carlo perturbations of each spectrum and subsequent re-fitting to find the 1$\sigma$ spread. An example of this fitting is shown in the left panel of Fig.~\ref{fig:grism_z} for one of the galaxies nearest the heart of the structure, galaxy 89. Our redshift of z=1.850$\pm$0.005 is consistent with the 3DHST-derived spectroscopic redshift for this galaxy (z=1.8491), and we find general consistency between our spectroscopic redshifts and the z$_{best}$ of 3DHST for our spectroscopic members.

We therefore identify this overdensity as a galaxy group at z=1.85, with six spectroscopically confirmed and approximately ten additional photometric members in this inner region. Photometric and spectroscopic redshifts are given in Table~\ref{tab:zs}. We consider galaxies contained within the smaller r=4.8" aperture, and the spectroscopically confirmed galaxy 88 just outside this aperture, as likely group members. We list 16 sources in total, combining sources present in the \citet{ref:M.Stefanon2017} catalogue, 3DHST catalogue and revealed by our NIRCam observations, as shown in Fig. \ref{fig:bcg} lower right (see Section~\ref{sec:sedgalaxiesmethods}).

\begin{table*}
\caption{IDs, SExtractor segmentation numbers, photometric and spectroscopic redshifts, stellar masses and SFRs of group member galaxies. z$_{\rm phot}$ are taken from the \citet{ref:M.Stefanon2017} catalogue, with errors derived from SED fitting in the FIR catalogue. SFRs derived in the FIR catalogue (SFR$_{\rm IR}$) and using SED fitting (SFR$_{FAST}$, this work) are labelled accordingly. Stellar masses are taken from our FAST++ fitting, with the 68\% errors shown. Some CANDELS IDs appear several times, associated with different segmentation IDs, derived from the higher-resolution imaging (see also Fig.~\ref{fig:bcg}). SFR$_{\rm IR}$ shown in bold indicate that the galaxies are detected at S/N$>$5 in the FIR regime, and SFR$_{\rm IR}$ are only shown for those galaxies with S/N$_{\rm FIR}>$2.5. 3$\sigma$ upper limits are shown on SFR values (using the errors on the derived SFR, from the FIR catalogue SED fitting or FAST SED fitting, respectively).}
\label{tab:zs}
\centering
\begin{tabular}{cccccccc}
\hline\hline
CANDELS ID & 3DHST & seg. & z$_{\rm phot}$ & z$_{\rm spec}$ & log$_{10}$(M$_{\star}$/M$_{\odot}$, FAST) & SFR$_{\rm IR}$ & SFR$_{FAST}$\\
& &  &  &  & & M$_{\odot}$/yr & M$_{\odot}$/yr \\
\hline
29816 & 38191 & 92 & 1.64$\pm$0.24 & 1.84$\pm$0.02 & 10.03$^{+0.21}_{-0.02}$ & <247 & <17.08 \\
30304 & 38274 & 78 & 1.72$\pm$0.18 & - & 10.50$^{+0.02}_{-0.03}$ & 108$\pm$32 & <8.58 \\
30147 & 38263 & 90 & 1.68$\pm$0.05 & 1.85$\pm$0.01 & 10.44$^{+0.05}_{-0.03}$ & <169 & <0.67 \\
29630 & 38497 & 88 & 1.95$\pm$0.27 & 1.86$\pm$0.01 & 8.80$^{+0.04}_{-0.05}$ & <146 & <0.44 \\
30186 & 38187 & 89 & 1.92$\pm$0.10 & 1.850$\pm$0.005 & 11.32$^{+0.05}_{-0.15}$ & \textbf{1810$\pm$219} & <168.53 \\
30186 & - & 99 & - & - & 10.14$^{+0.11}_{-0.02}$ & - & <2.85 \\
30336 & - & 85 & 1.78$\pm$0.21 & - & 10.45$^{+0.15}_{-0.07}$ & <262 & <27.03 \\
30186 & - & 100 & - & - & 9.64$^{+0.03}_{-0.11}$ & - & <0.64 \\
30144 & 38174 & 94 & 1.52$\pm$0.16 & 1.89$\pm$0.03 & 10.29$^{+0.02}_{-0.03}$ & \textbf{313$\pm$108} & <7.25 \\
29702 & 38356 & 87 & 1.64$\pm$0.11 & - & 9.83$^{+0.16}_{-0.02}$ & <276 & <4.50 \\
30329 & 38290 & 79 & 1.95$\pm$0.12 & - & 10.76$^{+0.11}_{-0.10}$ & 68$\pm$18 & <1.15 \\
28224 & 38527 & 83 & 1.53$\pm$0.23 & - & 8.88$^{+0.14}_{-0.13}$ & <120 & <1.45 \\
- & 38212 & 95 & - & - & 9.25$^{+0.03}_{-0.08}$ & - & <0.24 \\
- & - & 96 & - & - & 8.94$^{+0.04}_{-0.04}$ & - & <0.25 \\
- & - & 180 & - & - & 9.72$^{+0.07}_{-0.14}$ & - & <1.80 \\
- & 38531 & 181 & - & 1.85$\pm$0.05 & 8.53$^{+0.07}_{-0.07}$ & - & <0.12 \\

\hline
\end{tabular}
\end{table*}

\subsection{Photometric extraction for group galaxies}
\label{sec:sedgalaxiesmethods}
We began by identifying source extraction maps for our member galaxies and galaxies in the close projected surrounding of the group. A NIRCam F115W-F200W-F444W colour image of the group is shown in the lower left panel of Fig.~\ref{fig:bcg}. We used source extraction tool SExtractor \citep{ref:E.Bertin1996} to identify segments belonging to these galaxies, with the NIRCam F115W filter used as the detection image. We chose to use F115W as our detection image to allow for an appropriately high degree of segmentation in the complex central galaxy system and crowded field. Having done this, we used this segmentation map to extract photometry in all of our HST and NIRCam images, in order to derive consistent photometry for the same physical regions of galaxies. The resulting segmentation maps are shown in the lower right panel of Fig.~\ref{fig:bcg}. The flux density in each filter was taken to be the sum of flux densities in the pixels contained within each segmentation map, and the errors on these flux densities were also computed by SExtractor, using the CEERS RMS maps of the same part of the sky. We emphasise that the focus of our photometric detection and extraction was to select the candidate members present in the catalogues within the aperture defining the group, without selecting the diffuse light (see Section~\ref{sec:iclmethods}), rather than to perform blind detection of all sources.

Having measured photometry in this way, we derived the best-fitting template SEDs and star formation histories using the SED-fitting tool FAST++\footnote{https://github.com/cschreib/fastpp} \citep{ref:M.Kriek2009} applied to the HST and JWST data, setting a fixed redshift of z=1.85 for all members. We modelled the galaxies' star formation histories (SFHs) as (i) being exponentially declining with a characteristic timescale $\tau$ free to vary between log($\tau$)=7.5 and log($\tau$)=11, (ii) having a delayed-$\tau$ SFH, and also (iii) a pseudo-constant SFH (fixed log$_{10}$($\tau$/yrs)=11), all with the \citet{ref:D.Calzetti2000} dust extinction law, A$_{v}$ free to vary between 0 and 4 mag, and log(age/yr) between 7.5 and 10.1. We emphasise that this SED fitting was therefore more complex than simple stellar population modelling, which would most likely not be appropriate for these galaxies. Instead, the $\tau$- and delayed-$\tau$ models allow us to include a wide range of possible $\tau$ values, and therefore a range of stellar ages (as does the pseudo-constant SFH). Metallicity was fixed at Solar, 0.02. We found consistent results between exponentially declining and delayed-$\tau$ SFHs, and relatively poor fits with pseudo-constant SFHs. We therefore report the results of the exponentially declining SFH in Section~\ref{sec:results}.

\subsection{Photometric extraction for the intra-halo light}
\label{sec:iclmethods}
As seen in the lower left panel of Fig.~\ref{fig:bcg}, in addition to the light originating directly from galaxies, we also see compelling evidence for intra-halo light in the core of this forming structure, as further described below. We performed photometric extraction for the IHL, excluding the emission originating from galaxies. We did this by using SExtractor to create segmentation maps of the galaxies for all NIRCam bands - F115W, F150W, F200W, F270W, F356W, F410M, and F444W. We then combined all of the individual segmentation maps. To ensure that the extended, fainter outskirts of all galaxies are also included within this mask, we then radially extended our galaxy masks in all directions. The resulting `master mask', overlaid on each of the NIRCam and HST images used for photometric extraction of the IHL, is shown in Fig.~\ref{fig:cutouts}, where we show the group and the surrounding field. We show a large field of view in Fig.~\ref{fig:cutouts} to demonstrate the lack of bright halos around neighbouring objects, which could result from the PSF having very large wings, and thus mimic an IHL. We note that for the `peak' measurement, we deemed extending by 8 pixels (0.24" or 2.1~kpc) sufficient to balance masking the galaxy edges without losing too much of the high surface-brightness IHL in this region, whereas we use 12 pixels (0.36" or 3.1~kpc) when we extend our measurement to the full `core' region in the stacked image (see below). Defining an exact boundary between diffuse galaxy emission and the start of IHL is somewhat nuanced, and several different approaches have been taken in previous studies (e.g. fixed distance cutoffs from galaxy centres, investigation of IHL properties with increasing mask radius). We therefore created a number of galaxy masks, radially extending the original galaxy segmentation maps incrementally from 8 to 15 pixels, and verified that the exact number of pixels chosen does not have a significant influence on our results. Fig.~\ref{fig:cutouts} demonstrates that the master mask is sufficiently large to cover the full extent of galaxies, as shown by the lack of remaining emission, with the exception of the galaxy group. In the group region (the top left of the cutouts), we see bright diffuse emission between the group galaxies, already visible in NIRCam F115W and HST F125W, and becoming particularly prominent towards the longer wavelength NIRCam data.

\begin{figure*}
\begin{minipage}{\textwidth}
\includegraphics[width=0.5\textwidth, clip, trim=0.0cm 2cm 0cm 0cm]{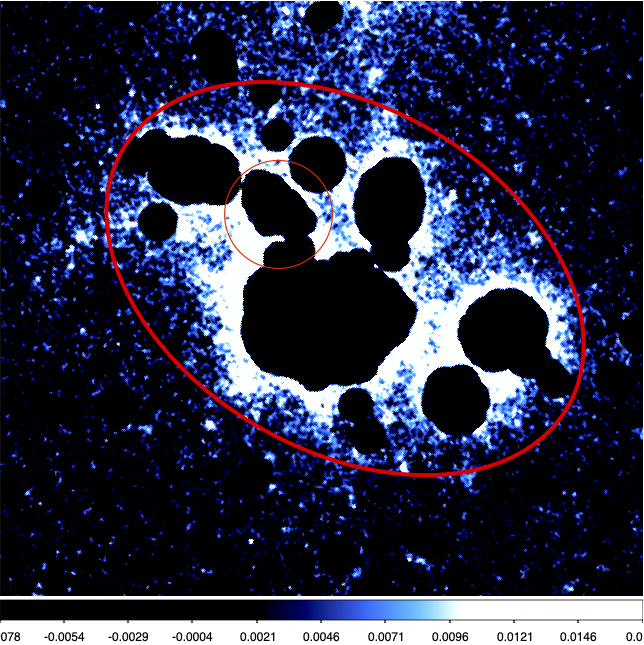}\hfill
\includegraphics[width=0.5\textwidth, clip, trim=12cm 6.65cm 13.4cm 7cm]{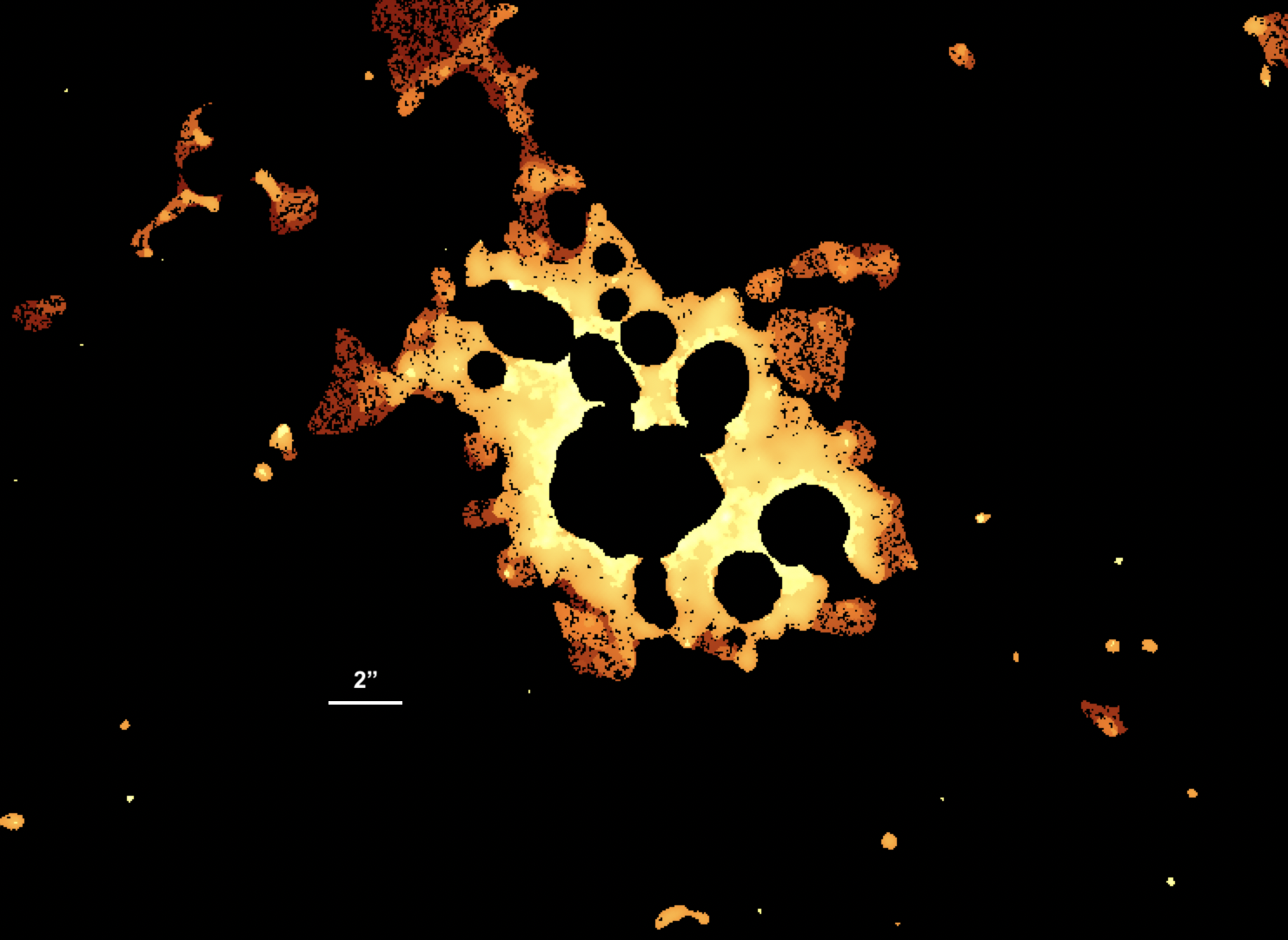}\hfill
\end{minipage}
\caption{Stacked images of the IHL. Left: S/N-weighted average, stacked image of the IHL in F356W, F410M and F444W, normalised to the F444W flux density, with 3 pixel Gaussian smoothing. The smaller red circle shows the aperture used for the peak measurement in the individual bands, and the larger red ellipse shows the aperture in which the core IHL F444W flux density was measured on this stacked image, in order to estimate the total IHL stellar mass in the core of the group. Right: adaptively smoothed stacked image.}
\label{fig:stacked}
\end{figure*}

We then performed aperture photometry on the background-subtracted images output by SExtractor at each wavelength, with the master mask placed over the galaxies (Fig.~\ref{fig:cutouts}). We inspected the images of the subtracted background itself, calculated by SExtractor, to verify that there was no local over-subtraction in the vicinity of the group, and that a smooth background is seen. If we take F444W, we see that the background images produced by SExtractor (\texttt{BACKPHOTO\_TYPE}=`GLOBAL') show a smoother background with less local substructure in the vicinity of the group than the CEERS v0.5 background image, and the median background over the group region from CEERS v0.5 is $\sim$5-12$\times$ higher than measured by SExtractor (the ratio between the two changes across different regions). In addition, the CEERS background images contain both positive and negative values, whereas our SExtractor images contain only positive background values. We place a small aperture of r=1.56" directly north of the central group BGG system, which we refer to as the peak IHL, shown by the small red circle in Fig.~\ref{fig:stacked} . The error on this aperture flux density was calculated by placing the same aperture at 1000 random empty positions over the surrounding field, and taking the standard deviation of aperture sums. We also performed photometry in annuli and apertures of larger radius around the peak region, but we find that the SED shape is not well constrained around the 4000\AA\ and Balmer breaks, so we do not attempt to derive SED shapes in these regions, and instead turn to stacking. The measured photometry for the IHL peak is given in Table~\ref{tab:photometry}.

\begin{table*}
\caption{Measured flux densities in the peak IHL aperture, in the given filters. The first three columns are HST, and the following are JWST/NIRCam.}
\label{tab:photometry}
\centering
\begin{tabular}{cccccccccc}
\hline\hline
F606W & F814W & F125W & F115W & F150W  & F200W  & F277W & F356W  & F410M   & F444W  \\
$\mu$Jy & $\mu$Jy & $\mu$Jy & $\mu$Jy & $\mu$Jy & $\mu$Jy & $\mu$Jy & $\mu$Jy & $\mu$Jy & $\mu$Jy  \\
 \hline
0.11$\pm$0.03 & 0.22$\pm$0.06 & 0.85$\pm$0.11 & 0.59$\pm$0.05 & 1.14$\pm$0.05& 1.23$\pm$0.06 & 1.36$\pm$0.08& 1.42$\pm$0.08 & 1.46$\pm$0.08 & 1.72$\pm$0.09\\
 \hline
 \end{tabular}
 \end{table*}
 
As with the galaxies themselves, we derived physical properties of the IHL using FAST++ SED fitting at fixed redshift, with the same parameter constraints as for the member galaxies. We modelled exponentially declining, delayed-$\tau$ and pseudo-constant SFHs, and found the results consistent between exponentially declining and delayed-$\tau$ templates. We report the exponentially declining results in Section~\ref{sec:results}. We show the best-fitting SEDs for the delayed-$\tau$ and pseudo-constant SFHs in Fig.~\ref{fig:sfh_comp} for comparison, where the pseudo-constant SFH is clearly ill-fitting. We also measured the flux density of the IHL regions using aperture photometry at 1.6$\mu$m in HST/WFC3 data, to compare with the F150W flux density measured from NIRCam. We found that the 1.5$\mu$m and 1.6$\mu$m flux density measurements are consistent with one another, so we continue with the deeper F150W data only for the SED fitting.

Finally, we created a stacked image of the IHL using the F356W-F410M-F444W filters, in order to gain the highest possible signal-to-noise ratio towards the fainter edges of the IHL distribution. Having measured the photometry in the peak region, we created an average image weighted using the signal-to-noise ratio (S/N) of the photometry in each filter, normalised to the F444W flux density. This is shown in Fig.~\ref{fig:stacked} (left). In the right panel, we show the same image but with adaptive smoothing, where we smoothed from scales of the PSF up to 3" diameter. Fig.~\ref{fig:stacked} highlights the morphology of the IHL in this inner part of the group, and we find a surface brightness 1$\sigma$ depth of 28.5 AB mag. over 1~arcsec$^{2}$, at rest-frame 1~$\mu$m in the stacked image. From this stacked image (left), we define a larger aperture, shown as the red ellipse, which we call the core of the group. We used this aperture to measure the flux density at F444W in this core region, and calculated a total stellar mass in the core by scaling the stellar mass measured in the peak region according to the ratio of the F444W flux densities in the peak and core regions. We assume the same mass-to-light ratio (M/L) in the peak and core regions, meaning the error on the core total stellar mass does not take into account possible differences in M/L. It is important to note that the core is much smaller than the virial radius of the group. Fig.~\ref{fig:stacked} demonstrates that here we are quantifying the IHL out to the 6.4"$\times$4.3" radius of the core aperture (55$\times$37~kpc at z~=~1.85), compared to an expected virial radius r$_{V}$$\sim$250~kpc \citep{ref:T.Goerdt2010}.

As seen in Fig.~\ref{fig:stacked}, a significant fraction of the sky within the apertures we use to measure IHL flux densities is masked, due to the need to mask the flux from galaxies. For this reason, when we estimate the stellar mass and star formation rate (SFR) within these regions, we increase the inferred stellar masses and SFRs by the ratio of masked to unmasked pixels. We therefore multiply the measured flux density by a factor of 1.69 in the peak circle, and 1.64 over the larger core ellipse, to account for the number of masked pixels, compared to the total number of pixels.

\begin{figure*}
\centering
\begin{minipage}{\textwidth}
    \includegraphics[width=0.5\textwidth, clip, trim=0.6cm 0.5cm 0.5cm 0.5cm]{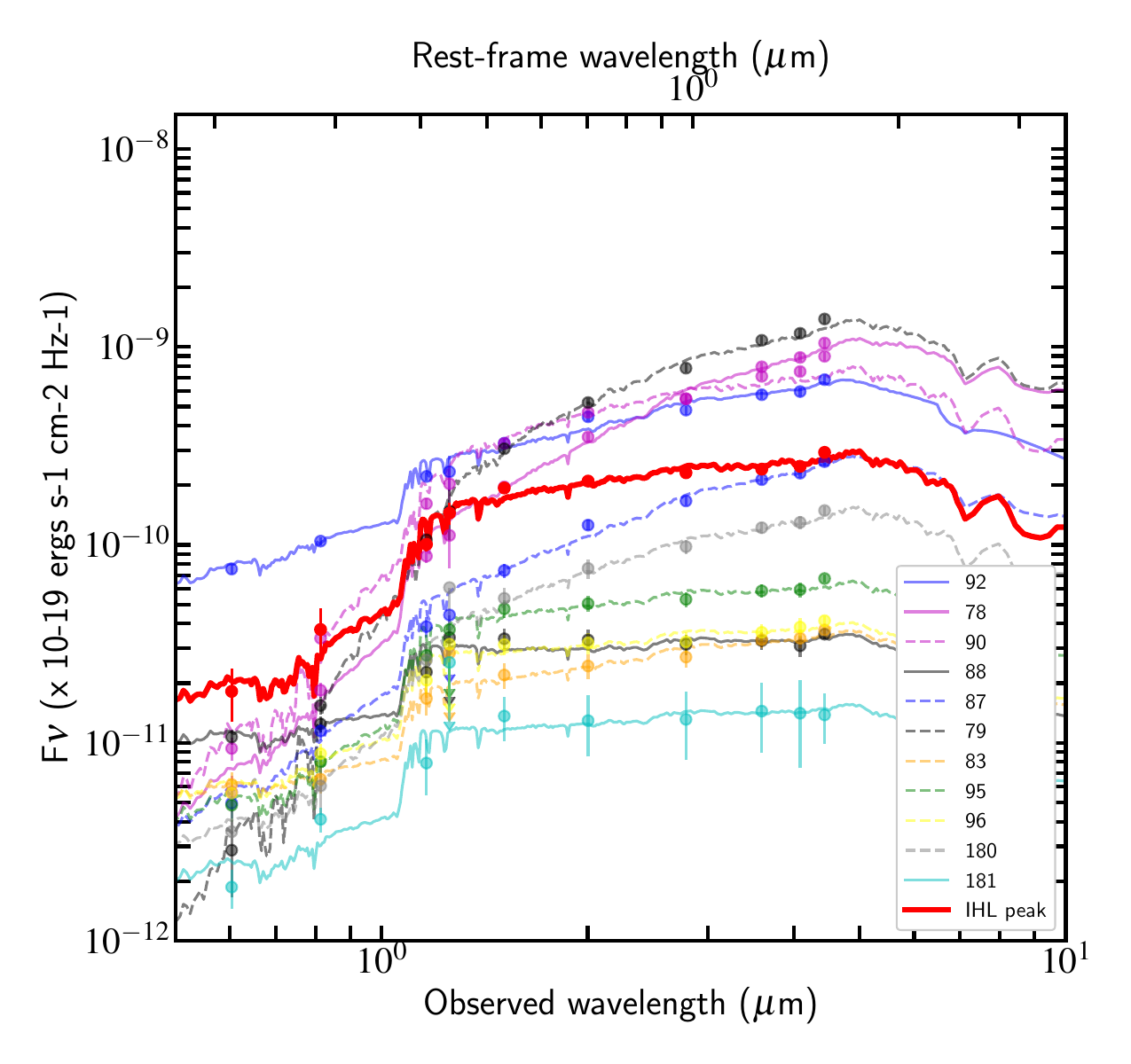}
      \includegraphics[width=0.5\textwidth, clip, trim=0.6cm 0.5cm 0.5cm 0.5cm]{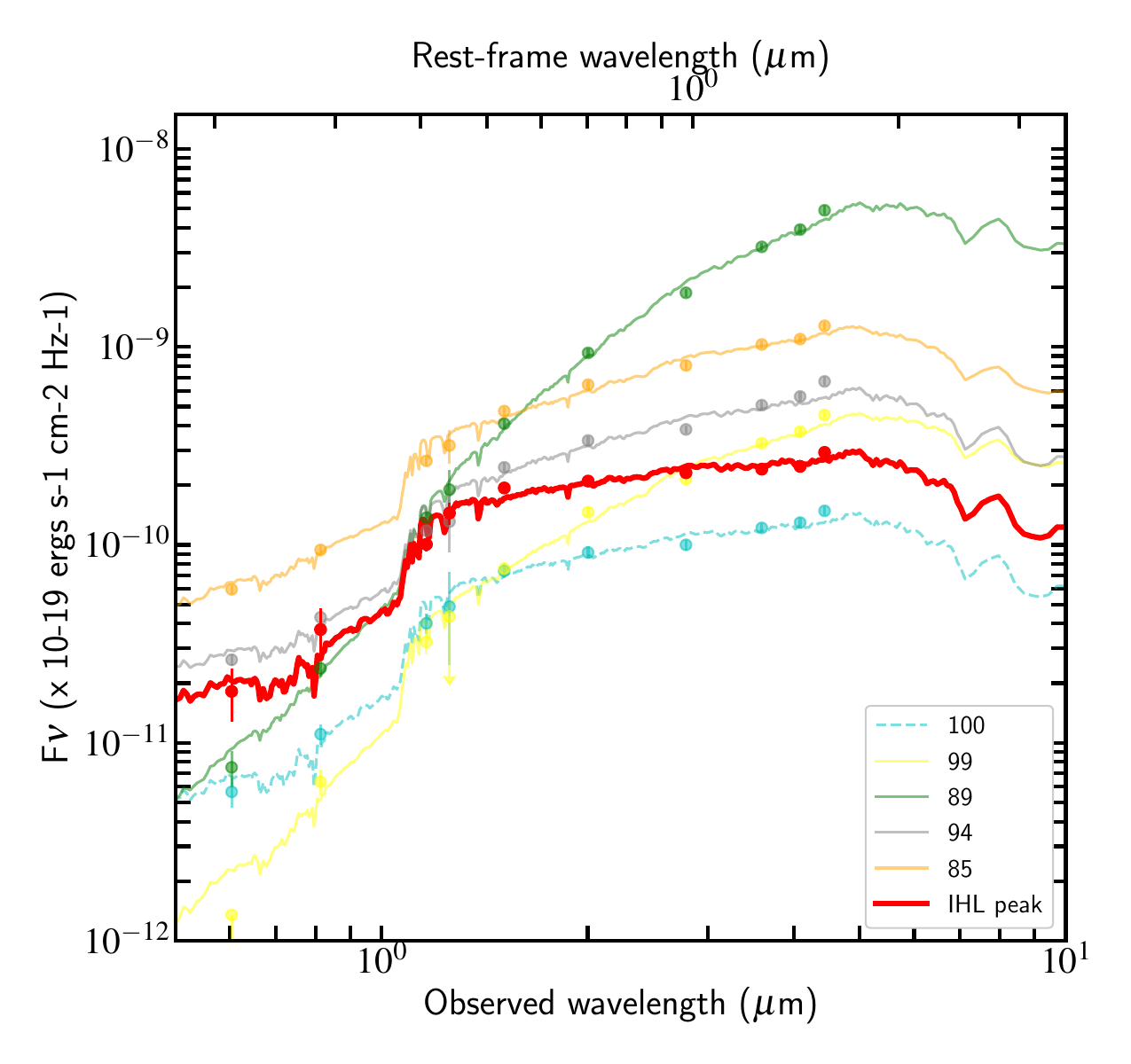}
    \end{minipage}    
\caption{Spectral energy distributions. Left: SEDs for galaxies, except for the BGG, in the group (various colours), in comparison to that for the IHL peak (red). Data points show photometric measurements, and correspondingly coloured lines show best-fitting templates to those data from FAST++. Right: As for the left panel, but showing only the peak IHL and the BGG galaxies. In both panels, the measured IHL flux densities have been increased according to the masked fraction in the photometry apertures, as discussed in the text.}
\label{fig:ICL_SED_SFH}
\end{figure*}

\subsection{Morphology measurements using GALFIT}

In order to quantify the morphology of the group galaxies, we used GALFIT 3.0.5 \citep{ref:C.Peng2002, ref:C.Peng2010}. As seen in Fig.~\ref{fig:bcg}, the morphology of the galaxies in the bluer bands is particularly complex (e.g. clumpy or disturbed, galaxy seg. IDs 78, 92). We are interested in using GALFIT to broadly quantify whether the galaxies are disk-like or spheroid-like, by measuring their S\'{e}rsic indices \citep{ref:J.Sersic1963}. For this reason, we used the F444W NIRCam images for the morphological fitting. We fixed the positions of galaxies to within a few pixels of the starting constraints, and we limited the effective radius to $\sim$15 pixels to ensure that the diffuse IHL does not unduly increase the sizes of the galaxies. Additionally, we masked (did not perform GALFIT on) the complex forming-BGG (seg. IDs 85, 89, 94, 99, 100), as we can judge visually that it is highly complex. We discuss the BGG in more detail in Section~\ref{disc:BCG}.


\begin{table*}
\caption{Summary of star-forming vs quiescent indicators for the group candidate members, with 68\% errors given. The UVJ column corresponds to the position of the galaxy in Fig.~\ref{fig:restcolours} (SF: star-forming, Q: quiescent, DSF: dusty star-forming).}
\label{tab:sfr_properties_comp}
\centering
\begin{tabular}{ccccccccccc}
\hline\hline
CANDELS ID & 3DHST & seg. & log$_{10}$(Age/yr) & log$_{10}$(tau/yr) & A$_{v}$ & t/$\tau$ & log$_{10}$(t$_{form,50}$/yr) & UVJ & n & $\chi^{2}_{red}$ \\
 \hline
29816 & 38191 & 92 & 8.10$^{+0.80}_{-0.00}$ & 7.50$^{+1.20}_{-0.00}$ & 0.86$^{+0.04}_{-0.42}$ & 3.98$^{+-0.00}_{-2.40}$ & 8.02$^{+0.74}_{-0.00}$ & SF & 2.3 & 0.93 \\
30304 & 38274 & 78 & 8.40$^{+0.10}_{-0.10}$ & 7.60$^{+0.20}_{-0.10}$ & 1.96$^{+0.14}_{-0.10}$ & 6.31$^{+1.63}_{-1.30}$ & 8.35$^{+0.09}_{-0.10}$ & Q/DSF & 1.1 & 0.81 \\
30147 & 38263 & 90 & 9.00$^{+0.20}_{-0.10}$ & 8.10$^{+0.30}_{-0.10}$ & 0.18$^{+0.26}_{-0.18}$ & 7.94$^{+0.00}_{-1.63}$ & 8.96$^{+0.19}_{-0.10}$ & Q & 2.7 & 1.76 \\
29630 & 38497 & 88 & 8.30$^{+0.20}_{-0.10}$ & 7.60$^{+0.40}_{-0.10}$ & 0.02$^{+0.12}_{-0.02}$ & 5.01$^{+-0.00}_{-1.85}$ & 8.23$^{+0.16}_{-0.10}$ & Q/SF & 0.5 & 0.33 \\
30186 & 38187 & 89 & 9.00$^{+0.20}_{-0.60}$ & 8.70$^{+0.10}_{-0.80}$ & 2.72$^{+0.38}_{-0.40}$ & 2.00$^{+1.17}_{-0.41}$ & 8.85$^{+0.22}_{-0.56}$ & DSF & - & 0.70 \\
30186 & - & 99 & 8.40$^{+0.60}_{-0.00}$ & 7.50$^{+0.80}_{-0.00}$ & 2.06$^{+0.10}_{-0.72}$ & 7.94$^{+2.06}_{-2.93}$ & 8.36$^{+0.58}_{-0.01}$ & Q/DSF & - & 1.41 \\
30186 & - & 100 & 8.90$^{+0.10}_{-0.40}$ & 8.20$^{+0.10}_{-0.70}$ & 0.22$^{+0.48}_{-0.22}$ & 5.01$^{+4.99}_{-0.00}$ & 8.83$^{+0.10}_{-0.39}$ & Q & - & 0.69 \\
 30336 & - & 85 & 8.50$^{+0.50}_{-0.30}$ & 8.00$^{+0.60}_{-0.50}$ & 1.06$^{+0.14}_{-0.36}$ & 3.16$^{+1.85}_{-0.65}$ & 8.40$^{+0.48}_{-0.26}$ & Q/SF & - & 0.70 \\
 30144 & 38174 & 94 & 9.00$^{+0.10}_{-0.10}$ & 8.50$^{+0.10}_{-0.10}$ & 0.58$^{+0.30}_{-0.18}$ & 3.16$^{+0.00}_{-0.65}$ & 8.90$^{+0.10}_{-0.12}$ & Q & 3.0 & 1.82 \\
 29702 & 38356 & 87 & 8.30$^{+0.70}_{-0.00}$ & 7.50$^{+1.00}_{-0.00}$ & 1.46$^{+0.12}_{-0.48}$ & 6.31$^{+0.00}_{-3.15}$ & 8.25$^{+0.65}_{-0.01}$ & Q/SF & 4.4 & 0.75 \\
 30329 & 38290 & 79 & 9.20$^{+0.30}_{-0.10}$ & 8.30$^{+0.30}_{-0.80}$ & 0.70$^{+0.18}_{-0.46}$ & 7.94$^{+31.87}_{-1.63}$ & 9.16$^{+0.30}_{-0.10}$ & Q & 4.0 & 0.58 \\
 28224 & 38527 & 83 & 8.60$^{+0.50}_{-0.50}$ & 8.20$^{+0.80}_{-0.70}$ & 0.42$^{+0.26}_{-0.42}$ & 2.51$^{+2.50}_{-1.51}$ & 8.48$^{+0.48}_{-0.46}$ & SF/Q & 3.0 & 0.22 \\
 - & 38212 & 95 & 8.70$^{+0.10}_{-0.20}$ & 7.90$^{+0.20}_{-0.40}$ & 0.06$^{+0.26}_{-0.06}$ & 6.31$^{+6.28}_{-1.30}$ & 8.65$^{+0.09}_{-0.19}$ & Q/SF & - & 0.36 \\
 - & - & 96 & 8.40$^{+0.20}_{-0.10}$ & 7.60$^{+0.30}_{-0.10}$ & 0.20$^{+0.20}_{-0.20}$ & 6.31$^{+1.63}_{-1.30}$ & 8.35$^{+0.19}_{-0.10}$ & Q/SF & - & 0.41 \\
- & - & 180 & 9.00$^{+0.30}_{-0.60}$ & 8.40$^{+0.30}_{-0.90}$ & 0.68$^{+0.62}_{-0.68}$ & 3.98$^{+6.02}_{-0.00}$ & 8.92$^{+0.32}_{-0.57}$ & Q & - & 0.16 \\
- & 38531 & 181 & 8.40$^{+0.20}_{-0.10}$ & 7.50$^{+0.40}_{-0.00}$ & 0.12$^{+0.36}_{-0.12}$ & 7.94$^{+0.00}_{-2.93}$ & 8.36$^{+0.18}_{-0.12}$ & Q/SF & - & 0.69 \\
\hline
\end{tabular}
\end{table*}

\begin{figure}
\centering
\includegraphics[width=0.5\textwidth, clip, trim=0.3cm 0cm 1.1cm 0.6cm]{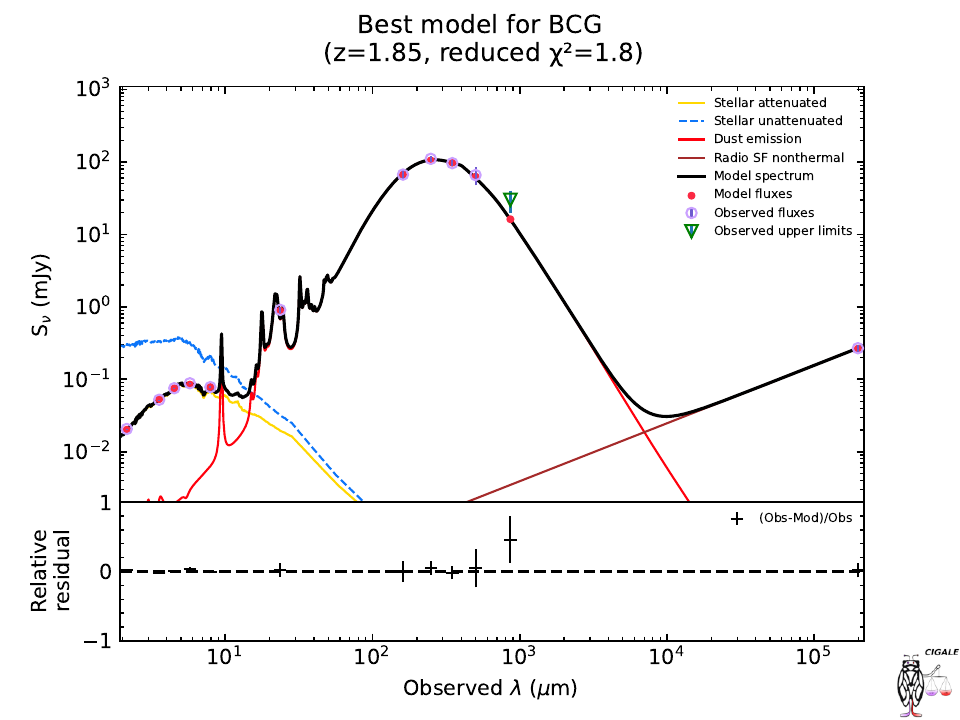}
\caption{FIR SED for the bright BGG seg. 89, from A. Le Bail et al. in prep., demonstrating no evidence for AGN components in the IR-radio spectrum. Several different components of the SED are shown, as labelled in the legend, based on CIGALE \citep{ref:M.Boquien2019} SED fitting. As discussed in the text, associations of Herschel fluxes with individual seg. maps in this work is not straightforward due to the large spatial resolution of Herschel, but we associate this Herschel flux to seg. 89 based on the super-deblended work of A. Le Bail et al. in prep.}
\label{fig:AGN}
\end{figure}

\begin{figure}[t]
\centering
\includegraphics[width=0.5\textwidth, clip, trim=0.7cm 0cm 0cm 0cm]{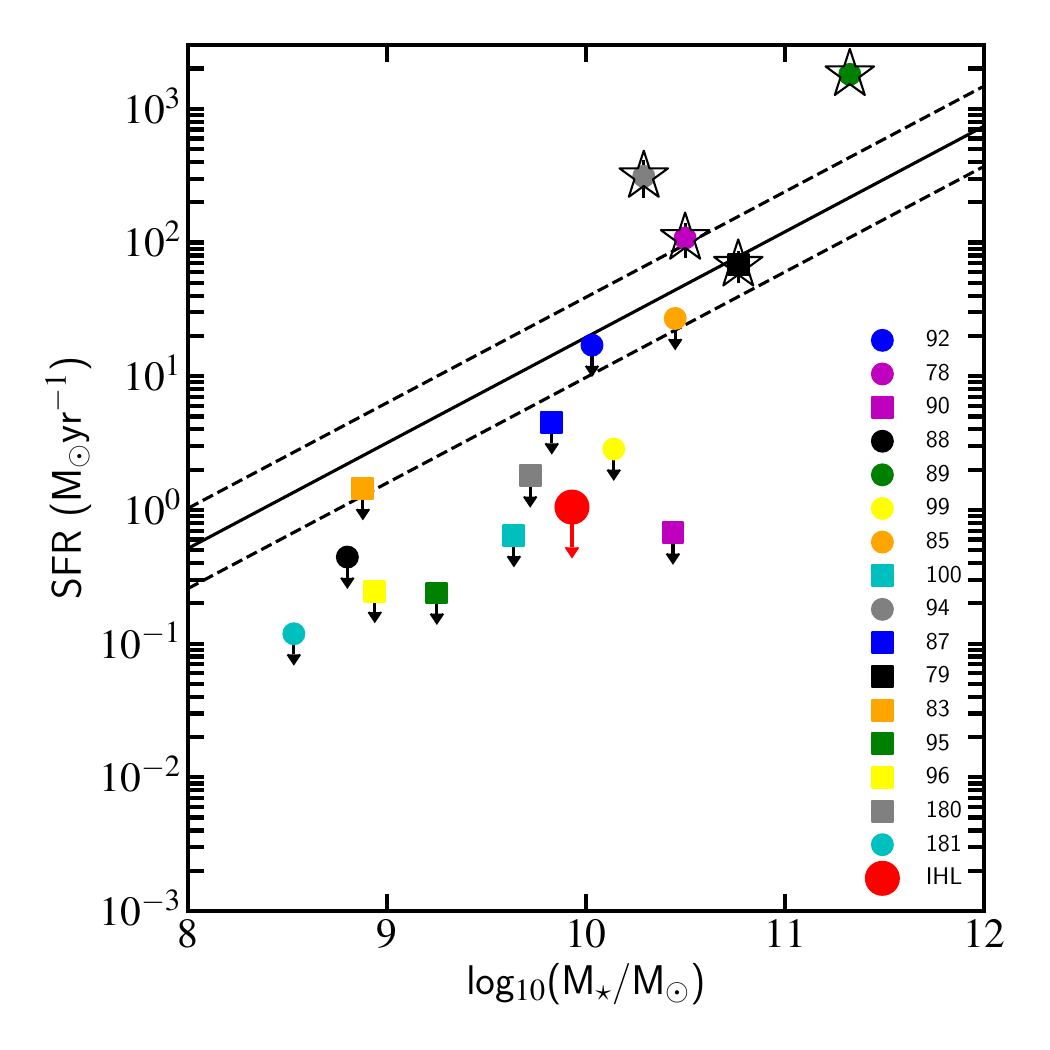}
\caption{Positions of the group galaxies with respect to the MS of star formation at z=1.85 (black solid line, \citealt{ref:M.Sargent2014}). The $\pm$0.3~dex scatter on the MS is shown, and the limits on SFR are at 3$\sigma$. The IHL peak is shown in red. SFRs have been taken from FAST++, with the exception of the 4 galaxies that have detected SFRs in the FIR catalogue (indicated with a star, see also Table~\ref{tab:zs}). 68\% errors are shown.}
\label{fig:MS}
\end{figure}

\section{Results and Discussion}
\label{sec:results}

\subsection{BGG formation}
\label{disc:BCG}
As seen in Fig.~\ref{fig:bcg}, there is clear evidence for the formation of the future Brightest Group Galaxy in the core of the galaxy group. We see several components in the high-resolution JWST imaging which are physically interacting with one another. This interaction is made clear by the very close projected separation of galaxies at the same redshift, and more interestingly, by the complex morphology of the combined proto-BGG structure. Although the irregular structure of the proto-BGG combined with the redshift information strongly suggests physical association between the components, we cannot rule out potential line-of-sight superpositions in the imaging data. We see variation in both visual morphology and colour, suggesting different levels of obscuration across the proto-BGG system, with two prominent bright and compact cores present in Fig.~\ref{fig:bcg}, which appear to be redder than the rest of the system. The core in the north-east of the BGG (seg. 89) is by far the most highly star-forming part of the group, with an obscured SFR estimated from the FIR catalogue of SFR = 1810$\pm$219~M$_{\odot}$/yr (Table~\ref{tab:zs}), based on super-deblended Herschel observations (we expect from the FIR-bright nature of this galaxy that SFR$_{tot}$$\sim$SFR$_{FIR}$, \citealt{ref:K.Whitaker2012}). There is also a somewhat tentative Herschel detection near to the second core, giving SFR = 313$\pm$108~M$_{\odot}$/yr (galaxy seg. 94/85), noting that the resolution of the Herschel data makes association between Herschel fluxes and specific segmentation maps from the NIRCam data in the BGG uncertain. Furthermore, the morphology of the bluer parts of the BGG (keeping in mind that Fig.~\ref{fig:bcg} lower left is an infrared image between 1.5-4.4~$\mu$m) surrounding these cores appears highly irregular, giving increased evidence to the fact that the component galaxies are actively interacting with one another. The diameter of the whole BGG system is approx. 3.6", a significant fraction of the r$\sim$5" aperture that defines this overdensity.

SExtractor identifies multiple separate regions within the BGG as shown in the lower right panel of Fig.~\ref{fig:bcg}, which are visually demarcated by both variation in colour and brightness in the lower left panel. We therefore consider 5 sources within the BGG - segmentation regions 85, 89, 94, 99, and 100. We show the SEDs of these galaxies, as well as the other group galaxies and the IHL, in Fig.~\ref{fig:ICL_SED_SFH}. The most noteworthy of these SEDs is that of the bright, star-forming core region 89, shown by the green solid line in Fig.~\ref{fig:ICL_SED_SFH} (right). We see that not only is the SED almost an order of magnitude brighter than the other member galaxies in the longest NIRCam filters, but the shape of the SED is significantly steeper, rising rapidly with wavelength up to around rest-frame 1.6-2.0~$\mu$m. Another of the BGG galaxies, seg. 99, also has a steeply rising (red) SED, shown by the solid yellow line in Fig.~\ref{fig:ICL_SED_SFH} right. As seen in Table~\ref{tab:sfr_properties_comp}, these two galaxies have the highest inferred dust extinction values (A$_{v}$~=~2.72$^{+0.38}_{-0.40}$ and 2.06$^{+0.10}_{-0.72}$respectively), explaining their redder SEDs. Galaxies 85, 94 and 100, the other elements of the BGG, are more moderately rising (bluer) in this wavelength range. When we compare with the rest of the group galaxies, we find that the BGG contains 69\% of the total stellar mass, totalling log$_{10}$(M$_{\star}/M_{\odot}$)=11.43 over the five components. The brightest and most star-forming component of the BGG, seg. 89, contributes 76\% of this BGG mass, being by far the most massive galaxy in the group at log$_{10}$(M$_{\star}$/M$_{\odot}$)=11.32$^{+0.05}_{-0.15}$ (Table~\ref{tab:zs}). This component of the BGG has also been detected as an X-ray active galactic nucleus (AGN) by \citet{ref:M.Brightman2014}, who identify this AGN as Compton-thick with a large column density log$_{10}$(N$_{H}$/cm$^{3}$)$\sim$24.6 using Chandra. We see from our SED fitting that the nucleus of seg. 89 also has very high dust attenuation, and it is therefore plausible that, at least in part, the AGN is obscured by the starbursting gas in this galaxy, and not only the torus \citep{ref:F.Bournaud2011, ref:A.Calabro2019}. However, we see no evidence of a mid-infrared-bright AGN component in our current data, nor a radio excess from the VLA 20~cm data (Le Bail et al. in prep.), that would affect our SED fitting, although this can be further confirmed with JWST/MIRI observations. We show the FIR SED-fitting of this galaxy in Fig.~\ref{fig:AGN}.

The highly infrared-luminous portions of the BGG are also interesting in terms of the group context itself. The majority of galaxies with such high IR-brightness, so-called ULIRGs or HyLIRGs, are massive galaxy mergers often found in the centre of galaxy groups or overdensities (e.g. \citealt{ref:R.Ivison2013, ref:H.Fu2013}). There is evidence for large scale associations around a significant fraction of intrinsic merger ULIRGs or HyLIRGs, meaning we may expect other dusty star-forming galaxies, extending out to 20-30 cMpc, that are likely part of the same structure at z=1.85. These galaxies would potentially give evidence of the massive cluster that this group will evolve into by by z=0. Investigating the presence of such galaxies, and whether such a large-scale structure is more compact than those at the same redshift in more extended fields, that do not show clear IHL in their central regions, would be valuable follow-up work.

\subsection{Spectral energy distributions and physical properties of group member galaxies}
\label{sec:sedresults}

We summarise the physical properties for group member galaxies in Tables~\ref{tab:zs} and \ref{tab:sfr_properties_comp}. SFRs and stellar masses are shown in Table~\ref{tab:zs}. S/N$_{\rm FIR}$ is defined as the combined S/N of a galaxy in `FIR' bands, taking into account Spitzer 24$\mu$m to VLA 20~cm data. SFRs in the FIR catalogue were then derived from super-deblended SED fitting by A. Le Bail et al. in prep., using IR-radio data, and are only shown for those galaxies that have S/N$_{\rm FIR}>$2.5, with those at  S/N$_{\rm FIR}>$5 shown in bold. We additionally show SFRs derived from our FAST++ SED fitting (see Section~\ref{sec:sedgalaxiesmethods}). We find no evidence for SFR at $>$3$\sigma$ in our group galaxies using the HST+NIRCam data in FAST++. It is interesting to note the differences in the two SFR estimates for our galaxies that are IR-detected. For those galaxies that have significant levels of star formation measured in the FIR-radio data, we see that the FAST++ SED fitting does not produce consistent results. This confirms that the star formation activity in these galaxies must be highly obscured. We also show stellar masses derived from our HST-JWST SED fitting in Table~\ref{tab:zs}, and we note that these are consistent with the stellar masses derived by \citet{ref:M.Stefanon2017}, taking into account nebular emission contamination.

In Fig.~\ref{fig:ICL_SED_SFH}, we compare the SED of the IHL (see Section~\ref{disc:ICL}) with the SEDs of the group member galaxies. We see that some galaxies have SED shapes that are much more steeply rising than that of the IHL - particularly the galaxies in the heart of the BGG (e.g. galaxy 89, green solid line, right panel). Some of the flattest SEDs we see in the NIRCam regime include the IHL (red), the faintest nearby group galaxy seg. 88, shown by the solid black line in Fig.~\ref{fig:ICL_SED_SFH} (left), as well as seg. 181 (cyan solid line, left panel). In addition to having a remarkably flat SED, seg. 88 also appears highly elongated in shape compared to the rest of the galaxies in the core of this group. In Fig.~\ref{fig:MS}, we show the positions of the group galaxies on the main sequence of star formation at z=1.85 (MS, \citealt{ref:D.Elbaz2007, ref:K.Noeske2007, ref:M.Sargent2014}). We see that the vast majority of the galaxies lie below the MS, with the only two members found above the scatter of the MS being members of the forming BGG.

The remainder of the galaxy properties derived from this SED fitting are shown in Table~\ref{tab:sfr_properties_comp}. We calculate t/$\tau$ values for the galaxies, and find that all have t/$\tau$$>$2, re-iterating their lack of star formation. In some cases, relatively low t/$\tau$ values suggest that the galaxies have only recently entered a quiescent state, whereas some galaxies have much higher values t/$\tau$$>$5, suggesting a more established quiescent state.

To get a more complete picture, we compare the t/$\tau$ values for the group galaxies with both their morphologies and their rest-frame UVJ colours. As also shown in Table~\ref{tab:sfr_properties_comp}, we find S\'{e}rsic indices n between 0.45-4.44 (for those galaxies not in the BGG), with the majority of galaxies having n$>$1, suggesting a more spheroid-like than disk-like morphology. We do not find a clear correlation between S\'{e}rsic index and t/$\tau$, but the majority of n$>$2.5 S\'{e}rsic indices agrees broadly with the quiescent-like t/$\tau$ values.

\begin{figure}
\centering
\includegraphics[width=0.45\textwidth, clip, trim=0.5cm 0.5cm 0.5cm 0.5cm]{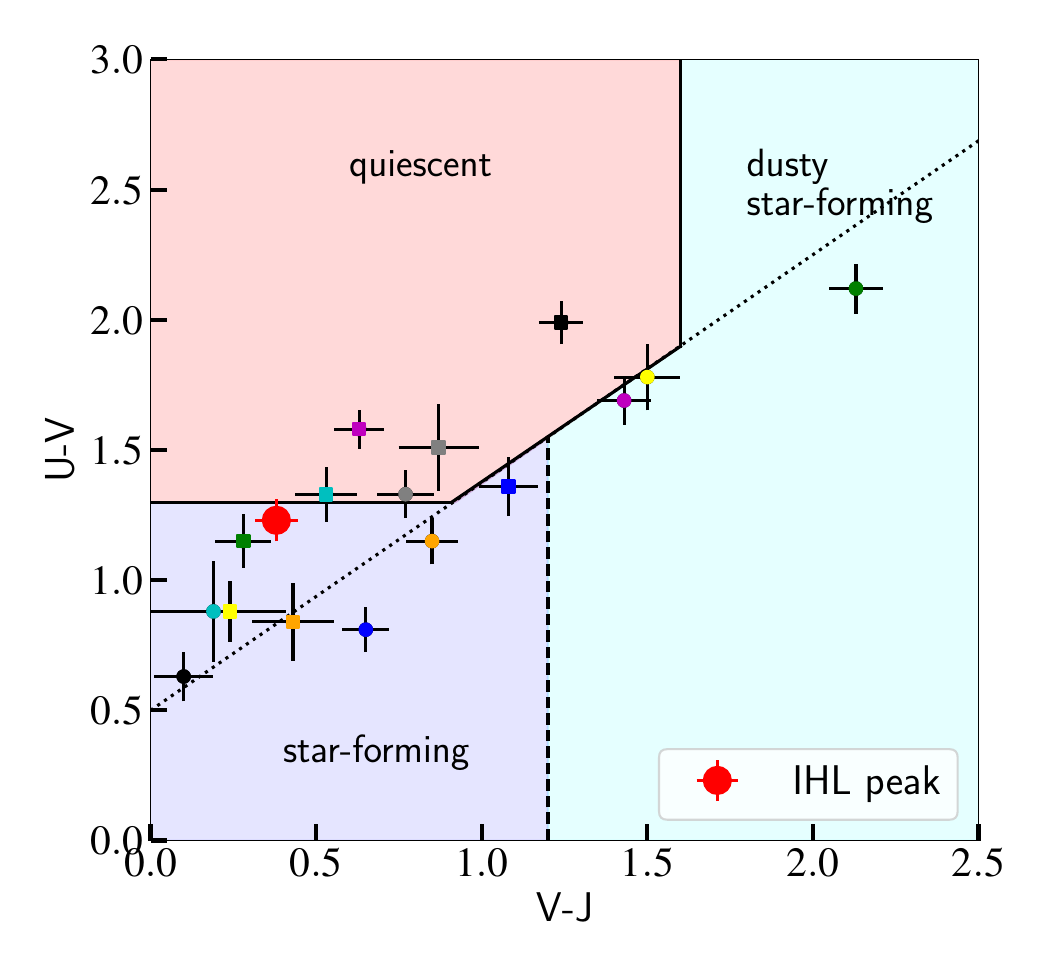}
\caption{UVJ diagram of the IHL (red) alongside the group galaxies (colours as in Fig.~\ref{fig:ICL_SED_SFH}). The dotted black line shows a proposed star-forming (lower) and quiescent (upper) division as discussed in \citet{ref:S.Belli2019} and \citet{ref:C.Deugenio2020}, for example. 68\% errors are shown.}
\label{fig:restcolours}
\end{figure}

In Fig.~\ref{fig:restcolours}, we show the UVJ diagram for the group galaxies, where the rest-frame magnitudes were calculated using FAST++. We note that we show two slightly different boundaries defining `quiescence' - one shown by the pink coloured section, with horizontal and vertical boundaries in U-V and V-J, respectively \citep{ref:R.Williams2009}, and the boundary shown by the dotted line, which continues in a straight line through UVJ space (e.g. \citealt{ref:S.Belli2019, ref:C.Deugenio2020}). As we might expect from the t/$\tau$ and morphology indicators, the group galaxies lie either close to the quiescent/star-forming dotted-line boundary, or further onto the quiescent side. If we were to instead take the coloured boundaries for quiescence, several galaxies would be consistent with star-forming colours, but again in the upper half closest to the quiescent boundary. The exception to this is the star-bursting BGG galaxy seg. 89, which is located in the dusty star-forming region on the right (green circle). This is consistent with the strong Herschel detection and corresponding high SFR. Putting all of this together, these three indicators (t/$\tau$, n, UVJ colours) suggest established or recent quiescence for the majority of the group galaxies, with the exception of the BGG. We summarise the UVJ regions occupied by the galaxies in Table~\ref{tab:sfr_properties_comp}, where `Q/SF' indicates that the colours would be consistent with quiescent or star-forming depending on the definition used, but we favour recent quiescence, as discussed.

Adopting the framework outlined in \citet{ref:R.vanDerBerg2014}, as was done in \citet{ref:E.Daddi2021}, we can estimate a total halo mass of the galaxy group from the stellar masses of the member galaxies. By studying z$\sim$1 clusters of galaxies, \citet{ref:R.vanDerBerg2014} find only small intrinsic scatter in the relationship between stellar and total halo mass, with little evolution in the range 0$<$z$<$1. Using their relationship log$_{10}$(M$_{200,\star}$/M$_{\odot}$)=12.44+0.59[log$_{10}$(M$_{200,halo}$/M$_{\odot}$)-14.5], we find a total halo mass log$_{10}$(M$_{200,halo}$/M$_{\odot}$)$\sim$13.1, assuming the relationship between stellar and halo mass does not evolve significantly from z=1 to z=1.85. We emphasise that our calculations are focused on the central 80~kpc of the group (within the r=$\sim$5" aperture). It is important to note that our methodology for detecting galaxy overdensities, discussed in Section~\ref{sec:methods}, naturally identifies the central peaks of the dark matter distributions in galaxy groups and clusters. As such, the full extent of our galaxy group is likely to lie beyond the radius r$\sim$5" we have selected. Bearing in mind that we are only taking into account galaxies within the central 80~kpc, there would be a correction factor of $\sim$2-3 to scale up to the total stellar mass within the virial radius, r$_{V}$$\sim$250~kpc \citep{ref:T.Goerdt2010}, assuming that the galaxy population follows an NFW profile, similar to high-z clusters \citep{ref:V.Strazzullo2013, ref:T.Wang2016}. This would then give log$_{10}$(M$_{200,halo}$/M$_{\odot}$)$\sim$13.6. If we instead use the \citet{ref:R.vanDerBerg2014} relationship between central BGG stellar mass and halo mass, log$_{10}$(M$_{BGG,\star}$/M$_{\odot}$)=11.66+0.42[log$_{10}$(M$_{200,halo}$/M$_{\odot}$)-14.5] to estimate halo mass, we find a total halo mass of log$_{10}$(M$_{200,halo}$/M$_{\odot}$)$\sim$13.6, using a BGG mass of log$_{10}$(M$_{BGG,\star}$/M$_{\odot}$)=11.3 from the single most massive component. We note that this relation flattens with increasing mass, and so becomes less reliable than for the total stellar mass scaling.

\begin{figure}
\centering
\includegraphics[width=0.5\textwidth, clip, trim=4cm 5.5cm 4cm 1cm]{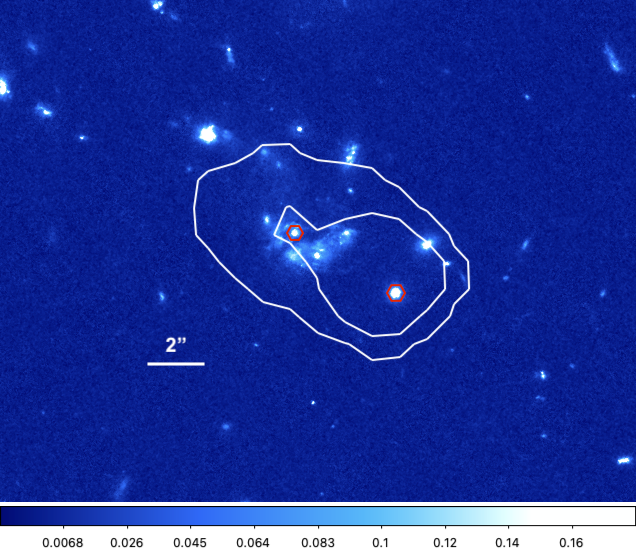}
\caption{Chandra 0.5-2~keV X-ray contours (white) overlaid on the NIRCam F115W image. The contours are at 2$\times$10$^{-15}$ ergs/s/cm$^{2}$/arcmin$^{2}$ and 2$\times$10$^{-14}$ ergs/s/cm$^{2}$/arcmin$^{2}$ \citep{ref:K.Nandra2015}. The two X-ray AGN are galaxies 89 and 90 (AEGIS 471 and AEGIS 470 respectively), shown by the red polygons.}
\label{fig:xray}
\end{figure}

We compare these halo mass estimates with that derived from X-ray constraints. As discussed in Section~\ref{disc:BCG}, a bright X-ray AGN is present in the BGG (seg. 89, AEGIS 471). A second X-ray AGN has also been identified, to the south-west of the BGG (seg. 90, AEGIS 470), which is unobscured and at lower X-ray luminosity \citep{ref:K.Nandra2015}. The X-ray contours in the core of the group are shown in Fig.~\ref{fig:xray}. Having removed the contributions of the two AGN point sources from the total emission, we find a 3$\sigma$ upper limit on the 0.5-2~keV X-ray flux of 4.4$\times$10$^{-16}$~ergs/s/cm$^{2}$ using a 12" aperture (and applying an aperture correction of 1.52), corresponding to an X-ray luminosity of L$_{X}$=3.2$\times$10$^{43}$~ergs/s. This implies an upper limit on halo mass of log$_{10}$(M$_{200,halo}$/M$_{\odot}$)~=~13.5 at z=1.85 \citep{ref:A.Leauthaud2010}. Combining the constraints from the stellar masses above (both total galaxy stellar mass and BGG stellar mass) and the X-ray limit, we adopt a best estimate of log$_{10}$(M$_{200,halo}$/M$_{\odot}$)$\sim$13.3 for the total halo mass within the virial radius of the group at z=1.85, with an uncertainty of $\sim$0.3-0.4~dex \citep{ref:E.Daddi2022a}. If we take the average evolution in M$_{200,halo}$ to be a factor $\sim$10 from z~=~2 to z~=~0, this will result in a Virgo-like halo mass for the group by z~=~0 \citep{ref:O.Fakhouri2010}.

Taking all of this information together, some interesting questions arise. For example, the most massive galaxies in the group appear to be the most actively star-forming, despite the fact we might expect the opposite. Despite the presence of emission lines in the 3DHST grism spectra of four galaxies outside of the BGG, there is little evidence for star formation activity in our data. The low SFRs in the lower-mass galaxies could be due to satellite quenching of galaxies towards the outskirts of the group, or inefficient gas-feeding processes. Perhaps we are capturing these low-mass galaxies in a transient, `green valley' regime, between star-forming and quiescent. However, the fact that we observe several galaxies with these low SFRs might instead suggest that this phase is more persistent, or long-lived. Quenching mechanisms such as RAM-pressure stripping lead to relatively slow quenching, and might therefore give rise to such observations. Given the above halo mass estimate, this group is therefore outside of the M$_{200,halo}$ range in which we would expect efficient cold-stream accretion onto the halo \citep{ref:A.Dekel2006, ref:E.Daddi2022a, ref:E.Daddi2022b}. For group-like halos above M$_{stream}\sim$2-3$\times$10$^{12}$~M$_{\odot}$, the maximum expected persistency of cold accretion is at the level of $\sim$100~M$_{\odot}$/yr \citep{ref:E.Daddi2022a}. The observed SFR of galaxy 89, SFR~=~1810~M$_{\odot}$/yr, is therefore a significant excess ($\times$10-20) above what we could expect for the fuelling of star formation, from cold gas accretion from the cosmic web. Another hypothesis for the presence of this significant star formation is therefore that the cold gas fuelling the high SFR in the BGG may originate from a cooling flow - the cooling of hot gas at the highest density peak of the group potential well. This may also provide an explanation for why the vast majority of the star formation appears to be taking place in one single galaxy within the BGG, close to the centre of the group \citep{ref:M.McDonald2018, ref:A.Fabian2022}. Alternatively, the SFR excess (compared to expectations from cold accretion) might be due to rapid and efficient consumption of residual gas in the group, as a result of the multiple mergers in the forming BGG. These hypotheses merit further investigation, in particular using deep X-ray observations or the Sunyaev-Zel'Dovich effect to explore the peak of the dark matter distribution (e.g. \citealt{ref:R.Gobat2018}).

\begin{table*}
\caption{FAST++-derived physical properties of the intra-halo light, with 95\% errors given both here and in the text. The stellar masses and SFRs in the peak (core) have been scaled up by a factor 1.69 (1.64), to account (linearly) for the masked regions. The M$_{\star}$ and SFR are conservative as the total IHL profile may be more peaked and/or extended than what we are assuming. The 3$\sigma$ upper limit is shown on SFR.}
\label{tab:iclphys_props}
\centering
\begin{tabular}{cccccccccc}
\hline\hline
ID & log(M$_{\star}$/M$_{\odot}$) & SFR & A$_{v}$ & log($\tau$/yr) & log(Age/yr) & t/$\tau$ & log(t$_{50}$/yr) & UVJ & $\chi^{2}_{red}$\\
 &  & M$_{\odot}$/yr & mag & & &  & &  &  \\
\hline
IHL Peak & 9.93$^{+0.02}_{-0.08}$ & $<$1.1 & 0.0$^{+0.3}_{-0.0}$ & 8.2$^{+0.2}_{-0.7}$ & 8.9$^{+0.1}_{-0.3}$ & 5.0$^{+10.8}_{-1.0}$ & 8.8$^{+0.1}_{-0.3}$ & Q & 0.8\\
IHL Core & 10.65$^{+0.02}_{-0.13}$ & & & & & & & & \\ 
\hline
\end{tabular}
\end{table*}

\subsection{Intra-halo light and group formation epoch}
\label{disc:ICL}

In addition to the member galaxies and BGG, the most noteworthy feature of Fig.~\ref{fig:bcg} is the intra-halo light that is revealed thanks to the sensitivity of NIRCam at these wavelengths. We see already by eye in Fig.~\ref{fig:bcg} (lower left) significant amounts of diffuse light extending to the north of the proto-BGG structure, at an intermediate colour between the reddest and bluest parts of the BGG. We also see this diffuse light clearly in Fig.~\ref{fig:cutouts}, where significant amounts of diffuse light are present throughout the core of the group, even after the galaxies themselves have been masked. The first hints of this IHL can already been seen in the HST/WFC3 F125W and JWST/NIRCam F115W images, although this becomes increasingly clearer in the redder NIRCam images. This material can be accounted for as the stellar material stripped as a result of interactions between galaxies during this active phase of structure build-up, and so gives an exciting insight into the formation history of this group. Moreover, the most prominent arc of the intra-halo light appears to extend north and east, connecting galaxies 92, 95, 96 and 180 to the BGG. This may suggest that one or more of these galaxies also previously experienced an interaction event with the BGG system.

\begin{figure}
\begin{overpic}[width=0.5\textwidth, clip, trim=0.6cm 0.5cm 0.5cm 0.5cm]{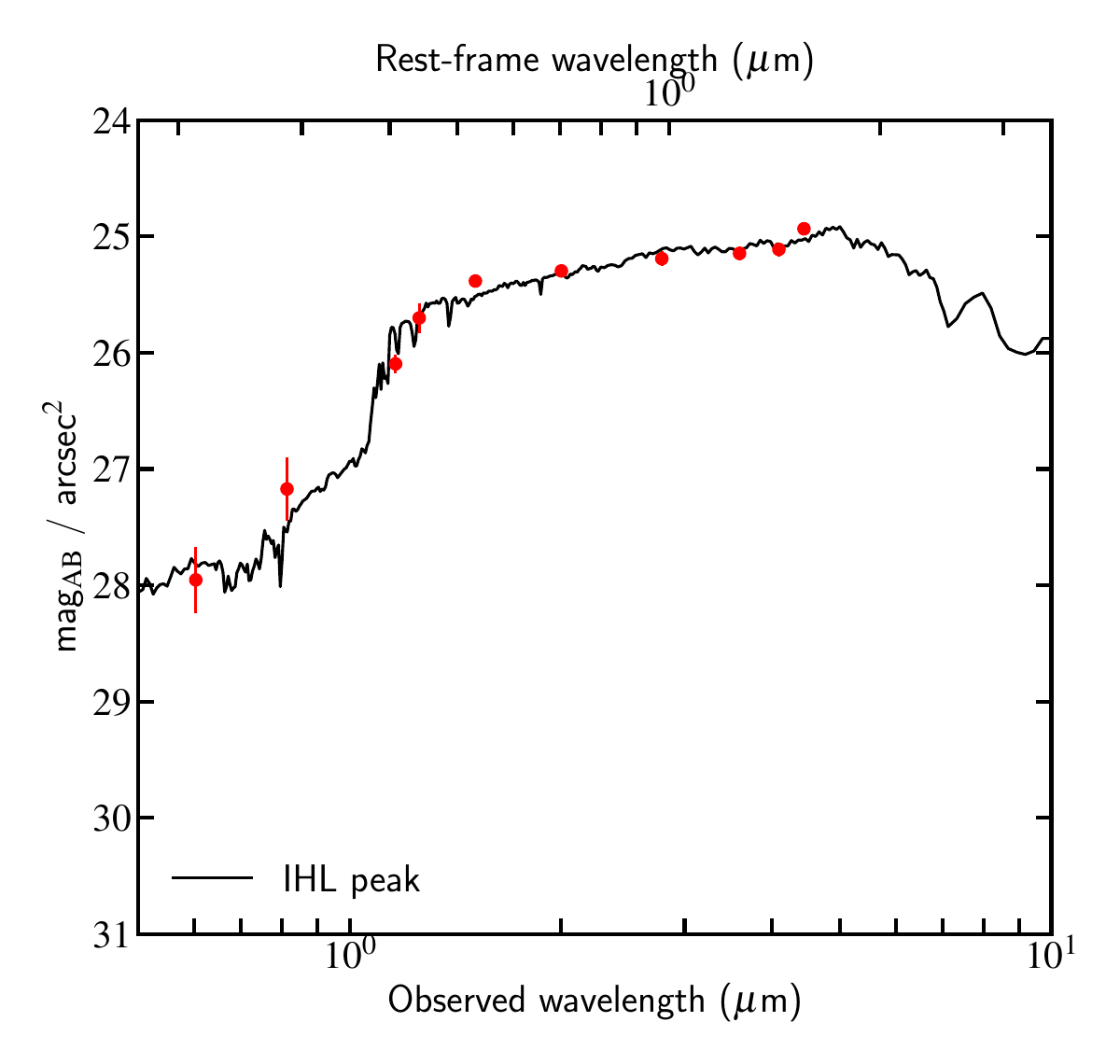}
     \put(33, 17){\includegraphics[width=0.3\textwidth, clip, trim=0.5cm 0.5cm 0.cm 0cm]{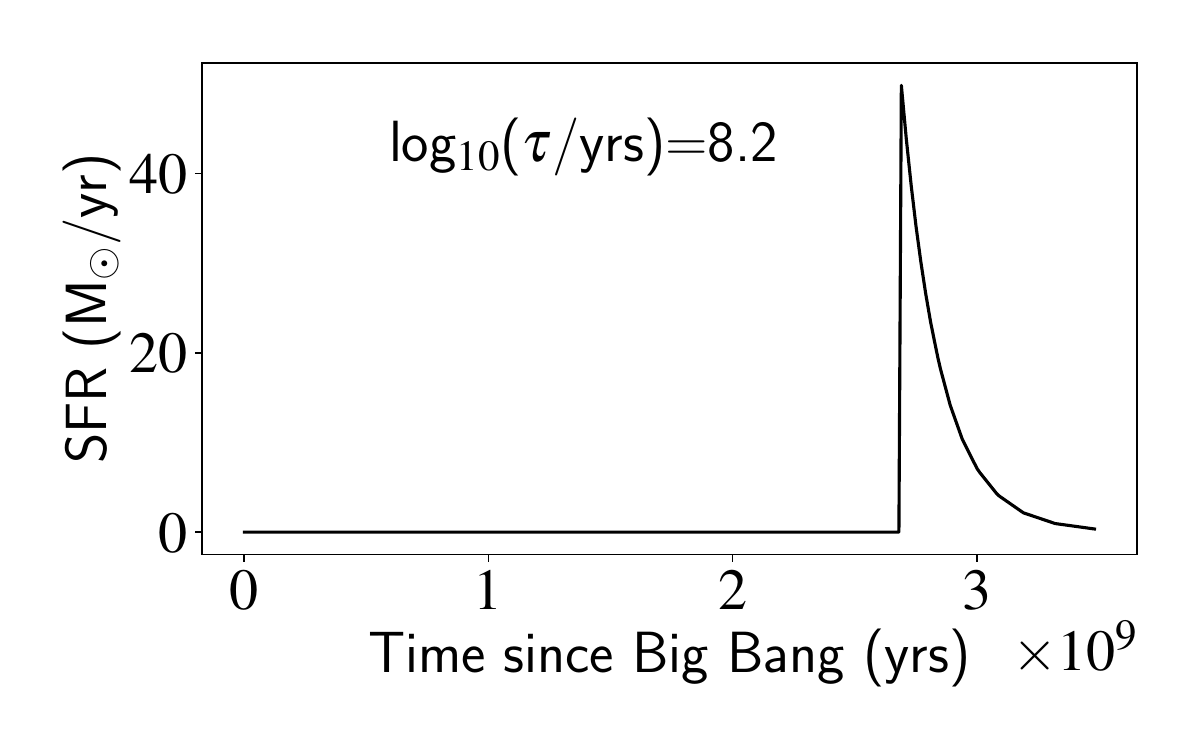}}
     \end{overpic}
\caption{Spectral energy distribution and star formation history for the IHL peak. Main panel: Surface brightness spectral energy distribution, where red shows observational data measured in a given aperture (HST and JWST), and the black line shows the best-fitting SED template found by FAST++. Detection threshold is taken at 2$\sigma$. Inset: The corresponding SFH for the peak region.}
\label{fig:ICL_SB}
\end{figure}

We show the surface brightness of the extracted IHL photometry, with the best-fitting exponentially declining SED overlaid, in Fig.~\ref{fig:ICL_SB}. We see a strong Balmer break, and the shape of the spectrum to wavelengths longer than this is fairly flat. In the inset panel, the star formation history of the peak region is shown. The x-axis represents the time since the Big Bang, increasing to the right until z=1.85.

We show the FAST++-derived properties of the IHL in Table~\ref{tab:iclphys_props}. We take the errors on these parameters as the 95\% range from $\chi^{2}$ minimisation, taking all of the solutions that give $\chi^{2}$ values within $\chi^{2}_{min}$+2.71 \citep{ref:Y.Avni1976}. We also show the distribution of parameter values versus $\chi^{2}$ of the fits in Fig.~\ref{fig:icldiagnostics}, as well as the relationship between A$_{v}$ and t/$\tau$ values.

We show the position of the IHL on the MS in Fig.~\ref{fig:MS} in red, and show the distribution of this position in Fig.~\ref{fig:icldiagnostics2}. The IHL is also shown on the UVJ diagram in relation to the galaxies in Fig.~\ref{fig:restcolours}, again in red. In agreement with the low SFR of the IHL derived from FAST++ (SFR=0.58$^{+0.52}_{-0.58}$~M$_{\odot}$/yr), and its relatively high t/$\tau$=5.0$^{+10.8}_{-1.0}$, the IHL UVJ colours shown in Fig.~\ref{fig:restcolours} suggest that the IHL borders the star-forming and quiescent regimes (with a preference towards quiescence, depending on the definition adopted).

We find that the IHL peak contains very little dust as judged by the amount of reddening - with A$_{v}$=0.0$^{+0.3}_{-0.0}$. This is noteworthy, as it may suggest that the mechanism leading to the formation of the IHL, stripping material from galaxies, was either not efficient at stripping the dust from the galaxies, or any dust stripped and placed into the IHL was very quickly destroyed in situ, for example by cosmic rays, or by X-rays, as often invoked to destroy dust in massive clusters. It also suggests that very little dust-enshrouded star formation can be taking place in the IHL. Comparing with the group galaxies themselves, we see that the galaxies have a fairly wide range of reddening values, between 0.02-2.72 (this highest A$_{v}$ being found in the dusty BGG galaxy 89). Several galaxies however also have A$_{v}$ consistent with 0, when the error is taken into account. If the dust content of the galaxies was already low when the IHL was formed, as might be the case if the galaxies were already quenching or if they had low stellar mass, this would also provide a possible explanation for the lack of dust reddening in the IHL. Alternatively, the different cross sections of the dust and stellar material may make it more likely that dust grains in the galaxies are left behind.

With this in mind, an important outcome of this fitting is the age of the stellar population in the IHL. FAST++ estimates both the `age' (stellar age of the best-fitting template for the SED, t) and the lookback time when 50\% of the stellar population was formed (t$_{50}$). In order to assess the formation epoch of the IHL, and that of the group galaxies, we consider t$_{50}$. As seen in Table~\ref{tab:iclphys_props}, log$_{10}$(t$_{50}$/yrs)~=~8.8$^{+0.1}_{-0.3}$. This gives a formation redshift, for 50\% of the stellar population, of z=2.27 (2.04-2.40 including the 95\% range). We see that some t$_{50}$ values of the BGG components (85, 99) are below that of the IHL, whereas some (89, 100, 94) are more comparable. In the total galaxy population, the minimum and maximum log$_{10}$(t$_{50}$/yrs) values are 8.02 and 9.16.

At least half the stars producing the IHL were therefore formed $\sim$0.6~Gyr before z=1.85, with a declining SFR characterised by timescale log($\tau$/yr)=8.2$^{+0.2}_{-0.7}$. In terms of age, t, compared to these $\tau$, we see that the age of the IHL is 5.0$^{+10.8}_{-1.0}$ times larger than the characteristic timescale. This suggests an evolved population that is not actively forming significant amounts of new, young stars. This is further supported by the low SFR and the very low A$_{\rm v}$ value that is consistent with 0, suggesting little-to-no dust, and therefore an absence of young, massive stars.

Finally, we find a mass of the IHL of log(M$_{\star}$/M$_{\odot}$)=10.65 in the larger, core IHL ellipse, by scaling the F444W flux density found in the peak aperture, to that found in the core aperture (5.3 times more flux is present in the core than in the peak, and we again take into account the masked pixels in the apertures). Combining this with the total stellar mass of the member galaxies listed in Table~\ref{tab:zs}, log(M$_{\star}$/M$_{\odot}$)=11.61, we find that the total stellar mass (in galaxies + in IHL) in the group's central region is log$_{10}$(M$_{\star}$/M$_{\odot}$)=11.66, meaning that the IHL contributes approximately $\sim$10\% of the total stellar mass in the group within the central R=80~kpc. This puts the contribution of the IHL at a comparable level to what is observed in local clusters (0$<$z$<$0.5, \citealt{ref:J.Krick2007, ref:C.Burke2015, ref:M.Montes2018, ref:Y.Jimenez2018}), although we keep in mind the caveat that some local studies present IHL light fractions rather than mass fractions (e.g. IHL light fraction as a function of halo mass, \citealt{ref:YT.Cheng2021}). This challenges some theoretical predictions based on simulations and semi-analytical modelling, which predict a steep decline in IHL mass fraction at z$>$1, becoming negligible by z$\sim$2 (e.g. \citealt{ref:C.Rudick2011, ref:E.Contini2014}). This IHL fraction instead supports those models that predict a less steep or negligible evolution (e.g. \citealt{ref:P.Behroozi2019}), and the observational findings of \citet{ref:H.Joo2023}. \citet{ref:C.Barfety2022} recently quantified the mass of diffuse stellar material in the z=1.7 SpARCS1049+56 cluster, having identified a massive, clumpy tail a few kpc from the Brightest Cluster Galaxy. They conclude that 15-21\% of the total IHL may be formed by a highly star-forming event in this structure, but we note that they find large amounts of molecular gas in the halo that is fuelling this star formation, which points to a very different picture to the dust-free IHL that we find for this group.

We note that both this IHL stellar mass and total galaxy stellar mass should be scaled up according to the full extent of the IHL profile and the total member galaxy population, respectively, to find the total over the entire group. We would however expect the profile of the IHL to follow the profile of the galaxy distribution \citep{ref:Navarro1996}. Additionally, thanks to the depth of the NIRCam data, we are able to probe to fairly low stellar masses (down to log$_{10}$(M$_{\star}$/M$_{\odot}$)=8.5), as demonstrated in Table~\ref{tab:zs}. Finally, if any of our non-spectroscopically confirmed galaxies are in fact interlopers and therefore not included in the galaxy stellar mass budget, this IHL mass fraction would increase.

We expect a dynamical timescale between galaxies of approximately 100~Myrs \citep{ref:diMatteo2007}, giving valuable information on the expected timescale of mergers and interactions in the forming BGG system. As this timescale is significantly shorter than the age of the IHL, it suggests that the merging BGG system is not the primary mechanism for creating the IHL. Furthermore, we can use the inferred halo mass to calculate a virial radius of the group of $\sim$250~kpc at z=1.85, from which we can estimate a rotation time at the virial radius of $\sim$3~Gyrs (representing the orbital crossing timescales for the group), for a velocity dispersion of $\sim$450 km/s \citep{ref:T.Goerdt2010}. The rotation time (orbital crossing timescale) at the radius we are focused on, R=80~kpc, would therefore be of the order 1~Gyr. This is therefore much more comparable with the age of the IHL than the dynamical timescale. Altogether, we see that the IHL in this galaxy group is likely not directly linked with the currently forming BGG - rather multiple interactions over a more prolonged timescale. Parts of the forming BGG are notably extended and infrared-luminous, particularly for this high redshift, at odds with the low dust content of the IHL. For galaxies 89 and 94, the high inferred dust attenuation supports the high FIR star formation rate in these parts of the BGG and their steeply rising SEDs. On the other hand, as seen in Table~\ref{tab:iclphys_props}, the multiple BGG components span a wide range of reddening values, with the non-star-forming components having significantly lower dust reddening, more consistent with what is seen in the IHL. Some studies up to z$\sim$2 claim to observe substantial dust gradients in galaxies, with the outskirts containing less dust than the galaxy centres (e.g. \citealt{ref:F.Liu2017, ref:S.Tacchella2018}). If material is preferentially stripped from the outskirts of galaxies to form the IHL, this may support our observation that the IHL has little dust, although there may be differential stripping between stellar material and dust. These dust gradients may indicate outside-in quenching, as was recently also suggested by ALMA imaging in galaxy group RO-1001 at z$\sim$2.9 \citep{ref:B.Kalita2022}. We should also keep in mind that the lack of young stars in the IHL does not mean that massive OB stars never entered the IHL. Rather, because of their short lifespans, these stars would die relatively quickly after being stripped, and as more material is added to the IHL, any young stars become a smaller fraction of the total stellar population. Finally, it is possible that RAM-pressure stripping in the galaxy group may have contributed to the creation of the IHL, although spectroscopic measurements would be needed to investigate this further.

\section{Conclusions}
\label{sec:conc}

We use JWST/NIRCam observations from CEERS to study the intra-halo light (IHL) and galaxy members in a forming galaxy group at z=1.85. We detected the group as a 5.3$\sigma$ overdensity of galaxies in the EGS field, and spectroscopically confirmed six of the 16 studied galaxy members, which go down to log$_{10}$(M$_{\star}$/M$_{\odot}$)=8.5.

We see clear evidence for a central, still-forming brightest group galaxy (BGG) in the NIRCam images, which is being assembled through mergers of several different galaxy components. We used multiple NIRCam filters, as well as archival HST data, to derive multi-band spectral energy distributions (SEDs) and corresponding physical properties for the BGG components and all of the member galaxies. The BGG components vary highly in star formation rate (SFR), morphology and colour, extending across $\sim$3.6" (30~kpc) of the sky. The BGG contributes 69\% of the total stellar mass in galaxies in the group, with one single massive member containing 76\% of the total BGG mass and a SFR$>$1810M$_{\odot}$/yr. This gives insight into massive galaxy build-up at this redshift where infrared-bright cores are surrounded by much bluer, more diffuse components. With the exception of three galaxies, two of which are in the BGG, we find that all galaxies in the group lie on or below the MS of star formation, indicating galaxies in the process of quenching, as confirmed by their position on the UVJ diagram. The galaxies display a range of S\'{e}rsic indicies, with a prevalence for high values, in line with a more evolved population.

We detect the intra-halo light in this galaxy group, in both HST and several JWST/NIRCam bands, allowing us to derive high-quality multi-band SEDs for the IHL for the first time at this high redshift. The IHL is extremely dust poor, being consistent with having no dust reddening (A$_{v}$=0.0$^{+0.3}_{-0.0}$), and a very low SFR=0.58$^{+0.52}_{-0.58}$~M$_{\odot}$/yr. The extremely low dust content of the IHL suggests that either dust grains were not efficiently stripped from galaxies during the formation of the IHL, that they were rapidly destroyed in the IHL, or that the galaxies involved already had a very low dust content, particularly in the outer regions most likely to be stripped. Having fit successfully with an exponentially declining star formation history, we find that t/$\tau$ for the IHL is t/$\tau$~=~5.0$^{+10.8}_{-1.0}$, suggesting the IHL is quiescent or recently quenched, consistent with the other indicators, including the rest-frame colours. The IHL contains an evolved stellar population, log$_{10}$(t$_{50}$/yrs)~=~8.8$^{+0.1}_{-0.3}$, giving a formation epoch for 50\% of the stellar material 0.6~Gyr before z=1.85, at z$\sim$2.27. In the core region of this group at z=1.85, we find that the IHL contributes $\sim$10\% of the total stellar material, comparable with what is observed in local clusters. This challenges expectations of IHL formation during the assembly of high-redshift clusters from previous theoretical modelling and simulations, and is more consistent with recently emerging studies predicting (or observing) significantly less evolution of IHL fraction with redshift. Through these studies of the IHL at wavelengths beyond what was possible with HST, JWST unveils a new side of group formation at intermediate-high redshift, for halo masses which will evolve into Virgo-like structures in the local Universe. In order to further understand this active group, measurements of the molecular gas content (with e.g. NOEMA), as well as spectroscopic information on the dynamics of the material in the galaxies, are crucial next steps.

\begin{acknowledgements}
We thank the referee for their comments which improved the clarity of this paper. RTC was supported by a postdoctoral fellowship from the Centre National d'\'{E}tudes Spatiales. MF acknowledges support from NSF grant AST-2009577
and NASA JWST GO programme 1727.
\end{acknowledgements}

%
%

\bibliographystyle{aa} 
\bibliography{/Users/rcoogan/Documents/Bibliography/EnvironmentNew}

\begin{appendix}

\section{Parameter versus $\chi^{2}$ distributions for IHL parameters from FAST++ fitting}
\begin{figure*}
\centering
\begin{minipage}{\textwidth}
\includegraphics[width=0.44\textwidth, clip, trim=0cm 0.cm 0cm 0.0cm]{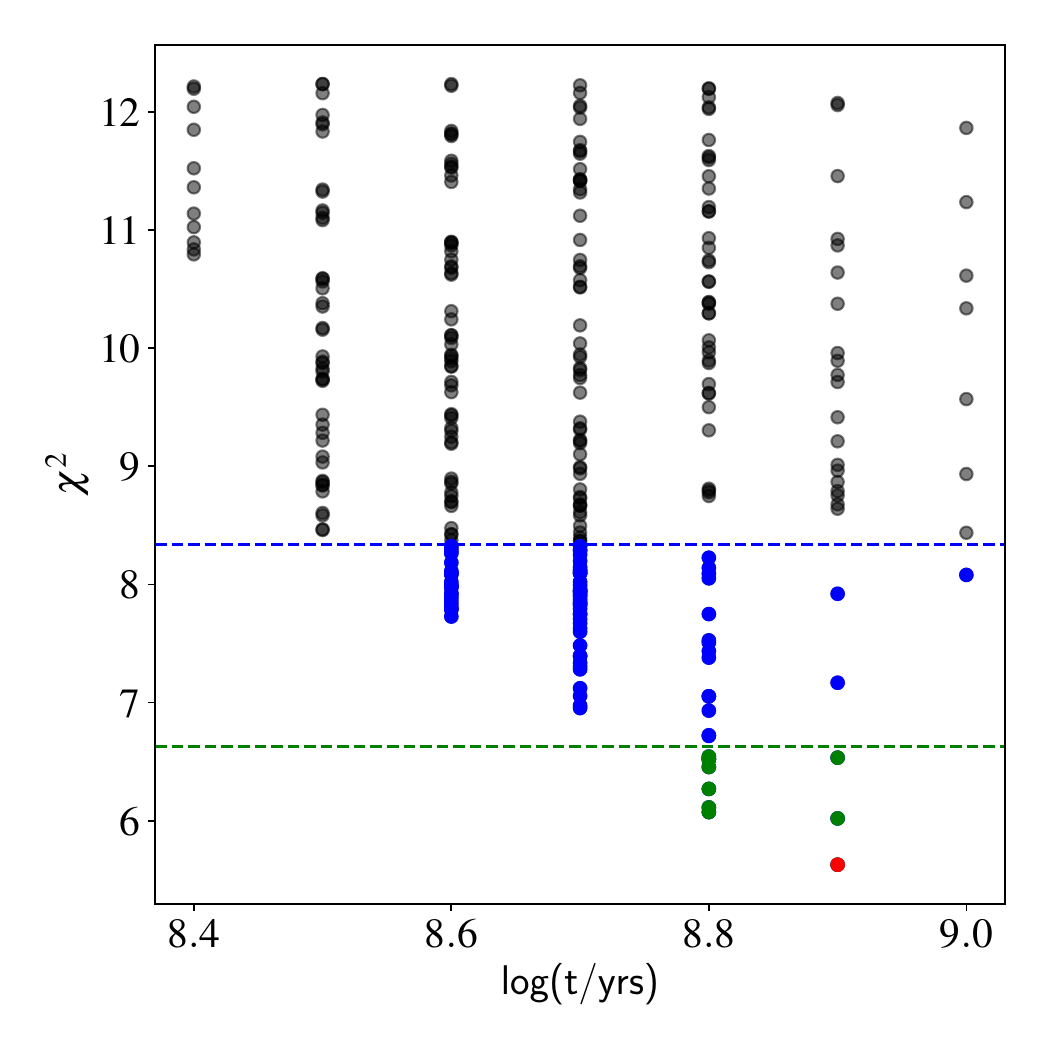}\hfill
\includegraphics[width=0.44\textwidth, clip, trim=0cm 0.0cm 0cm 0.0cm]{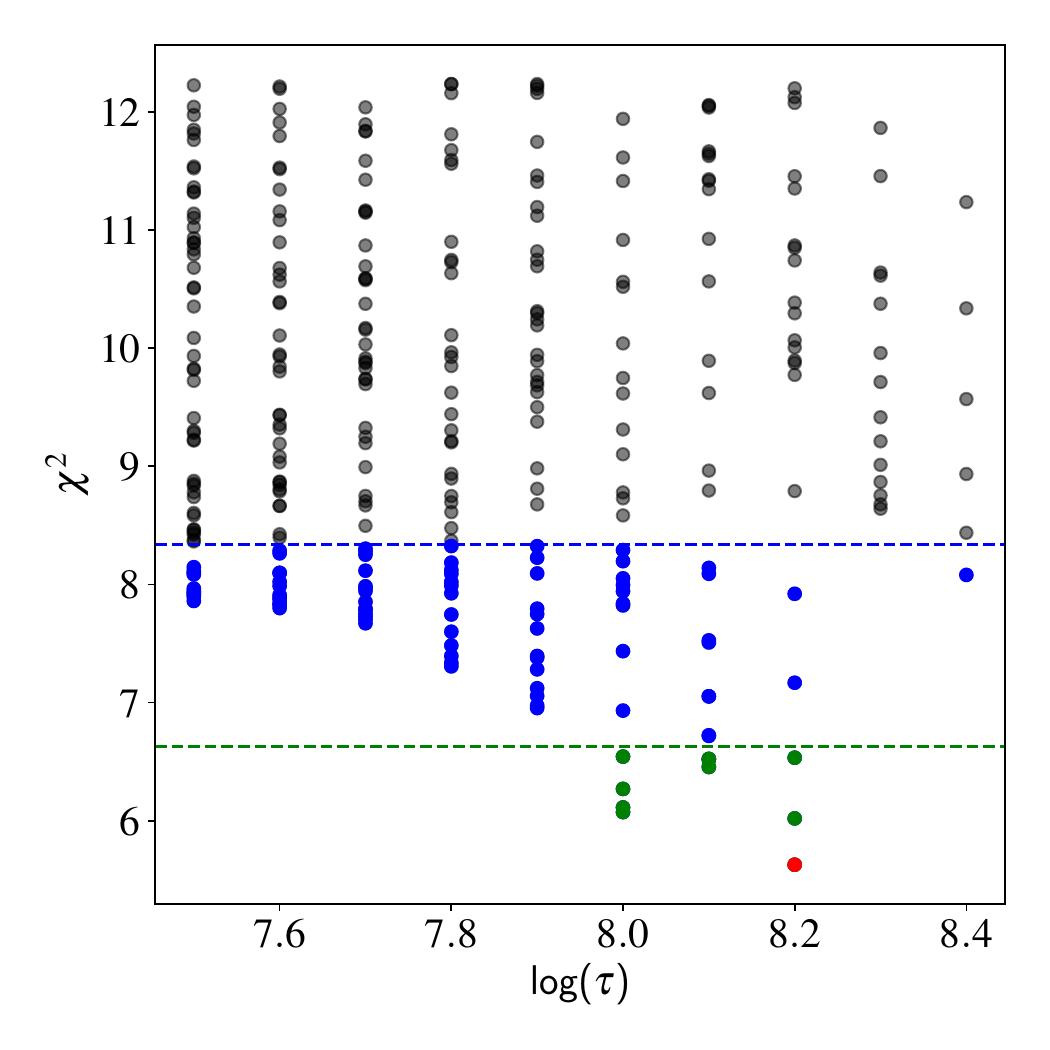}\hfill

\end{minipage}
\begin{minipage}{\textwidth}
\includegraphics[width=0.44\textwidth, clip, trim=0cm 0.0cm 0cm 0.0cm]{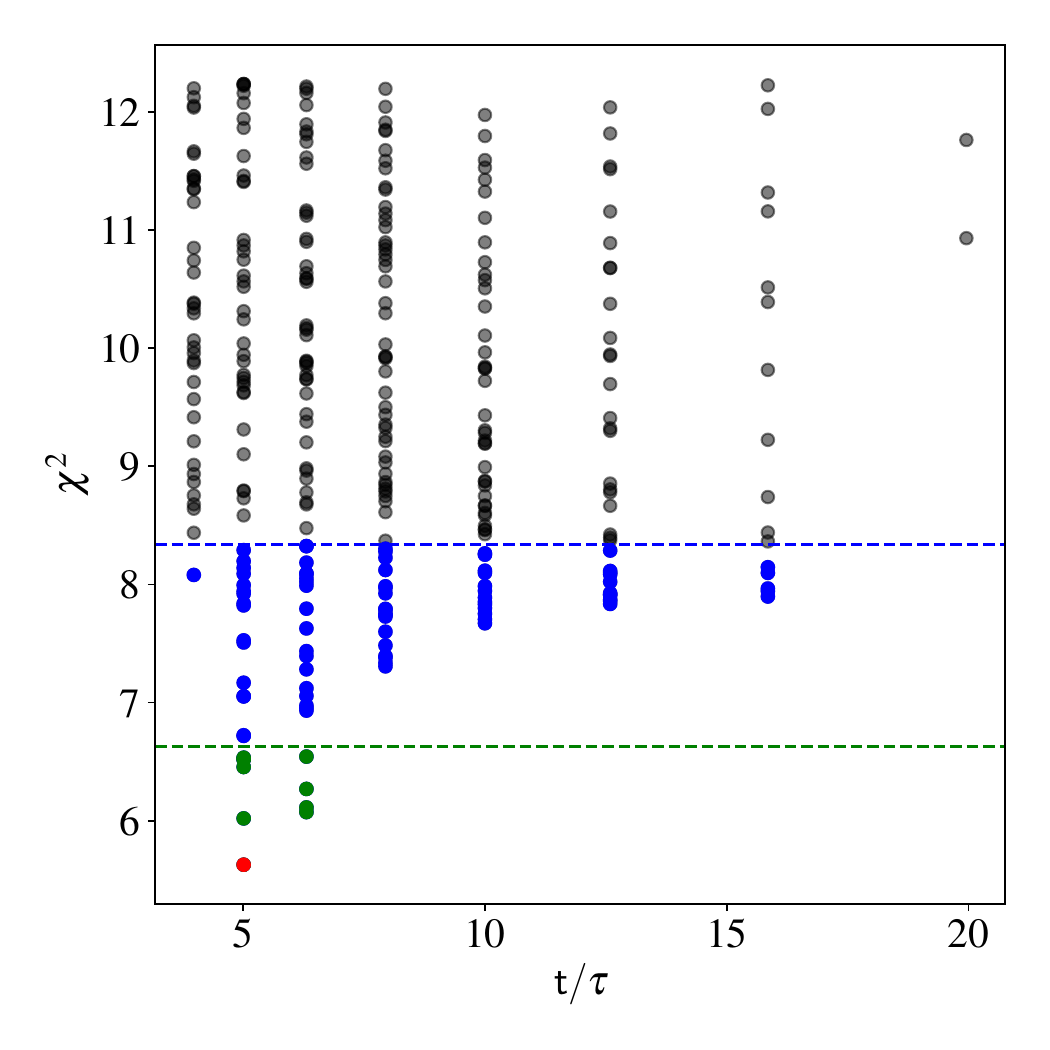}\hfill
\includegraphics[width=0.44\textwidth, clip, trim=0cm 0.0cm 0cm 0.0cm]{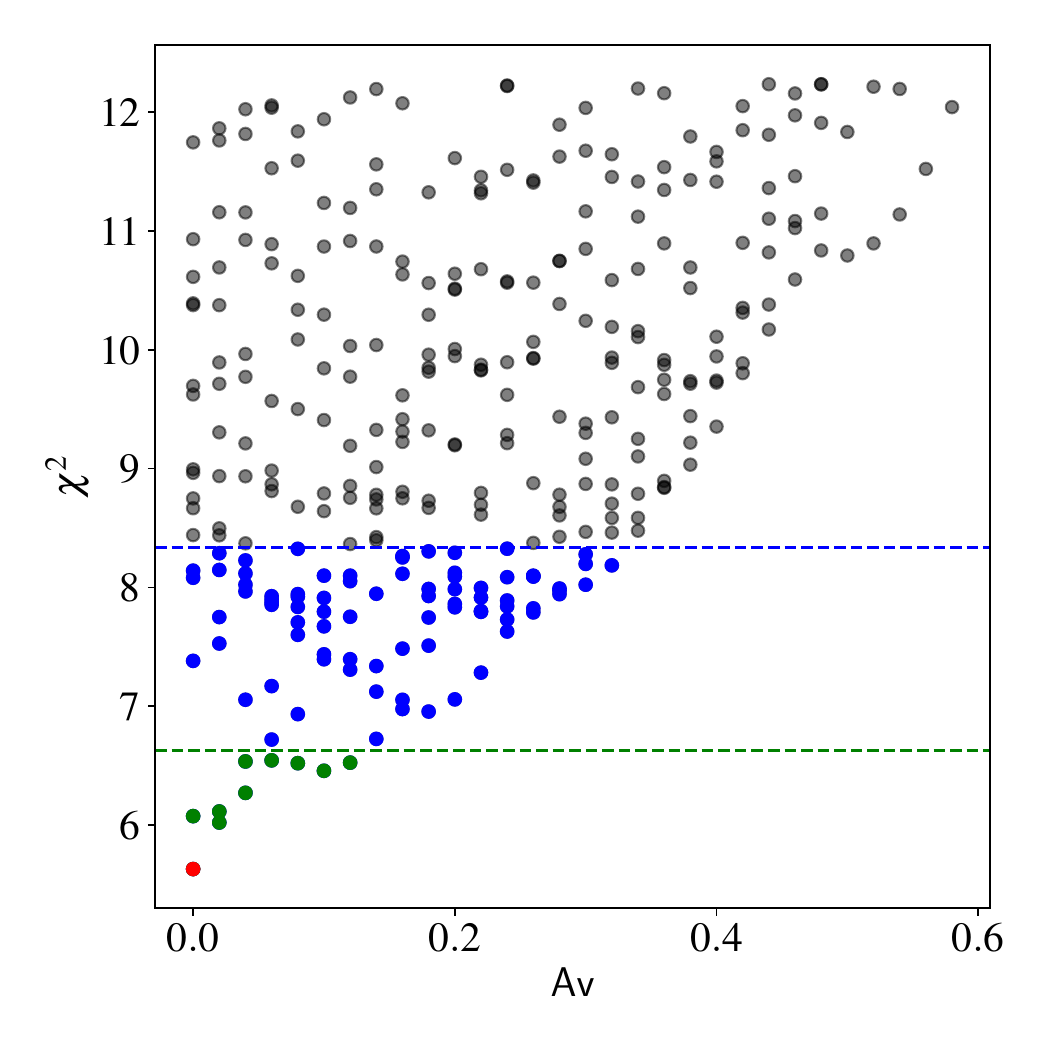}\hfill
\end{minipage}
\begin{minipage}{\textwidth}
\includegraphics[width=0.44\textwidth, clip, trim=0cm 0.0cm 0cm 0.0cm]{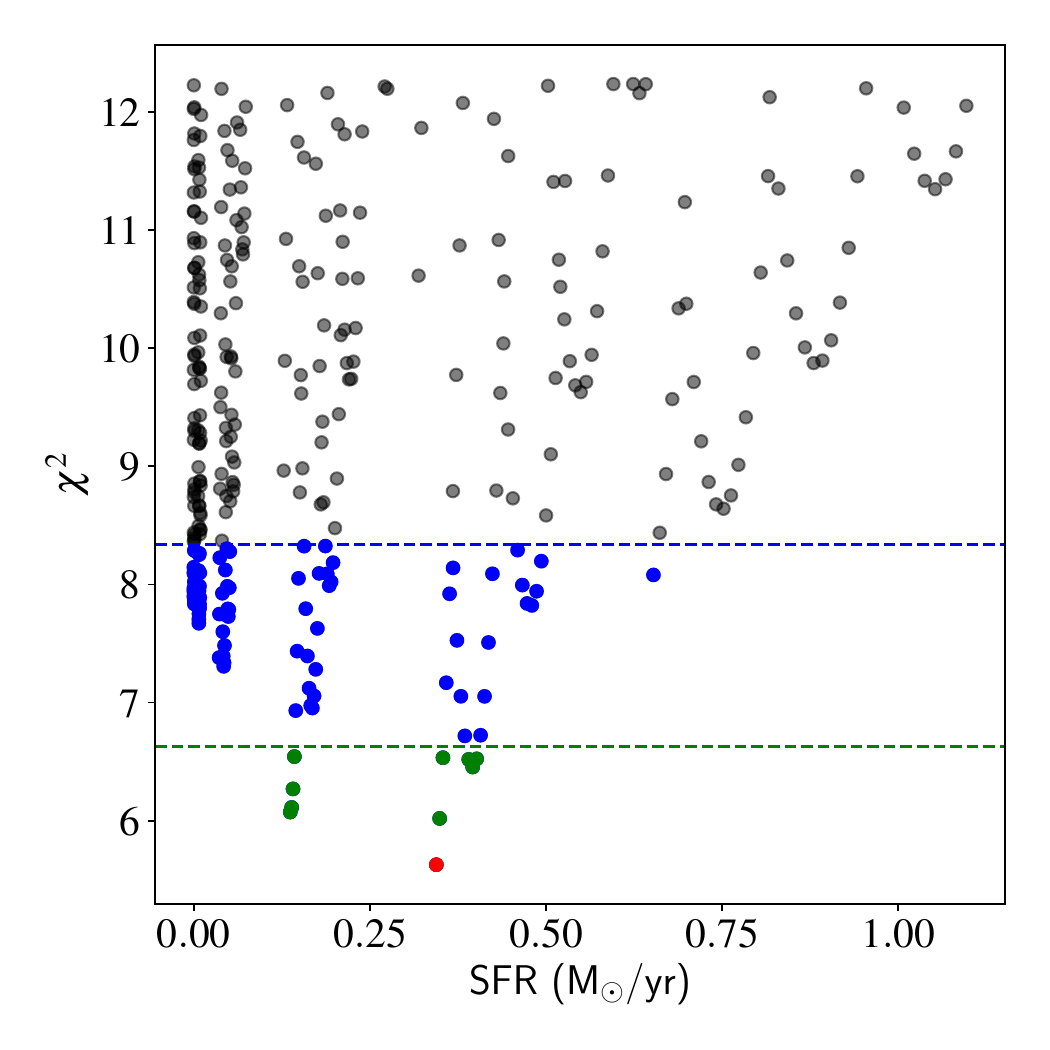}\hfill
\includegraphics[width=0.44\textwidth, clip, trim=0cm 0.0cm 0cm 0.0cm]{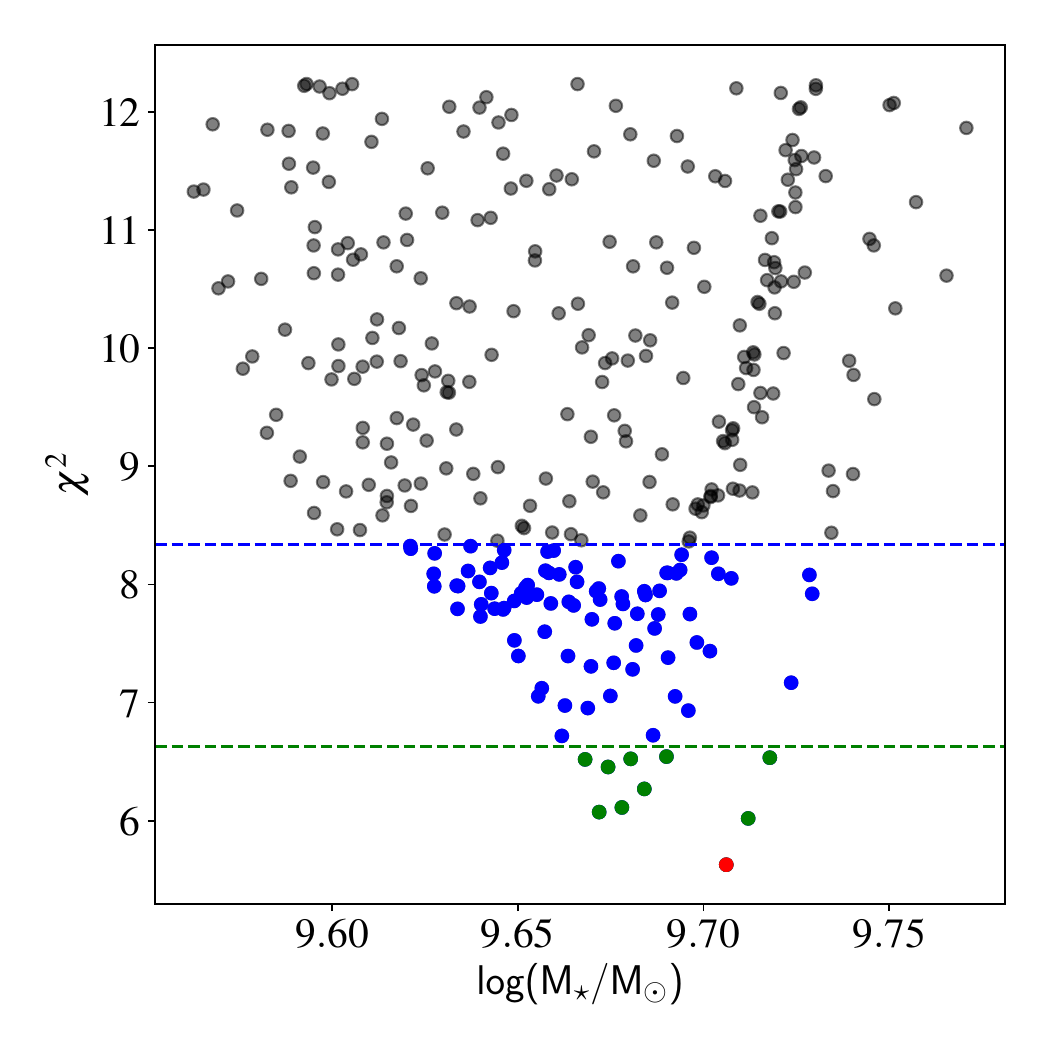}\hfill
\end{minipage}
\caption{Parameter vs $\chi^{2}$ distributions for the peak of the IHL, as labelled. Grey points are within 3$\sigma$, blue points are within 2$\sigma$, green are within 1$\sigma$. The red points show the values corresponding to the minimum $\chi^{2}$.}
\label{fig:icldiagnostics}
\end{figure*}

\begin{figure*}
\centering
\begin{minipage}{\textwidth}
\includegraphics[width=0.45\textwidth, clip, trim=0cm 0.0cm 0cm 0.0cm]{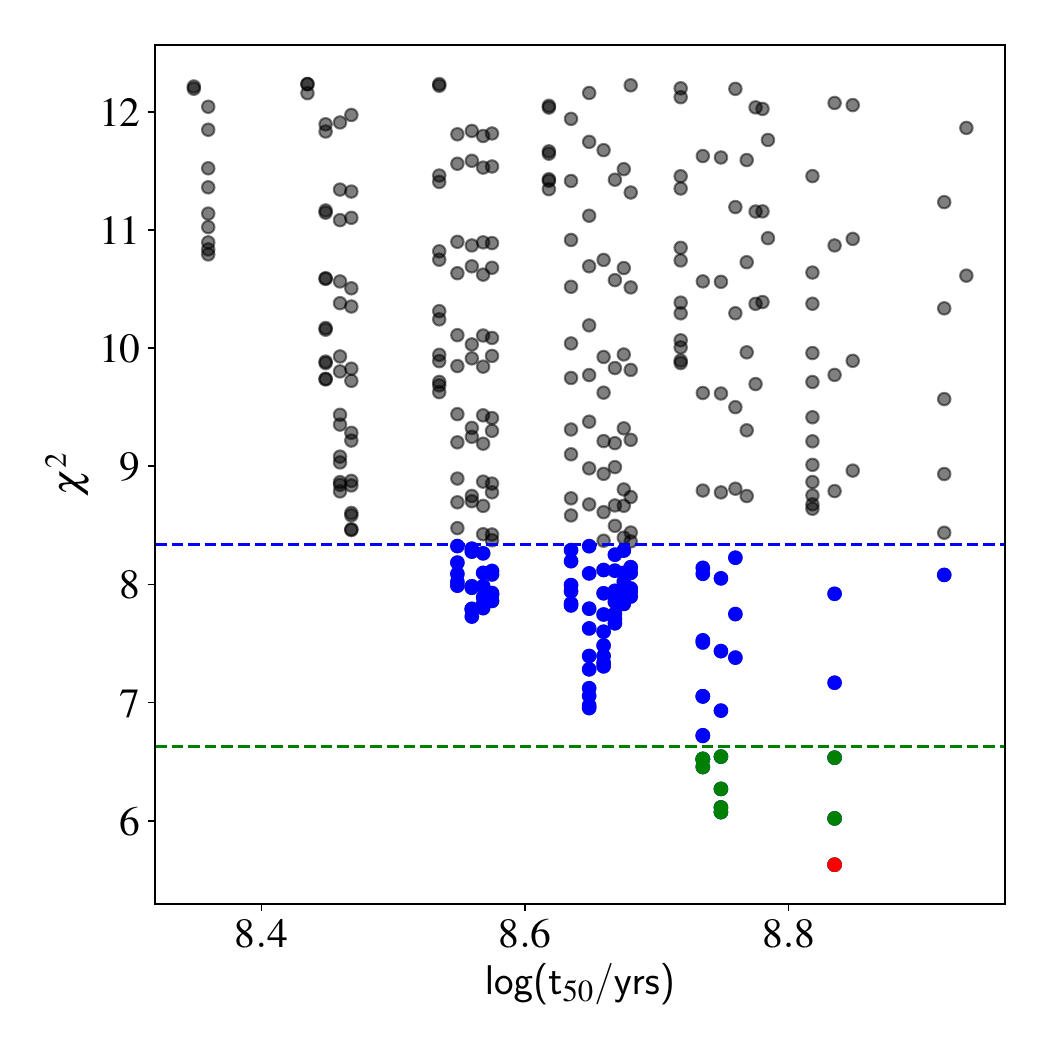}\hfill
\includegraphics[width=0.45\textwidth, clip, trim=0cm 0.0cm 0cm 0.0cm]{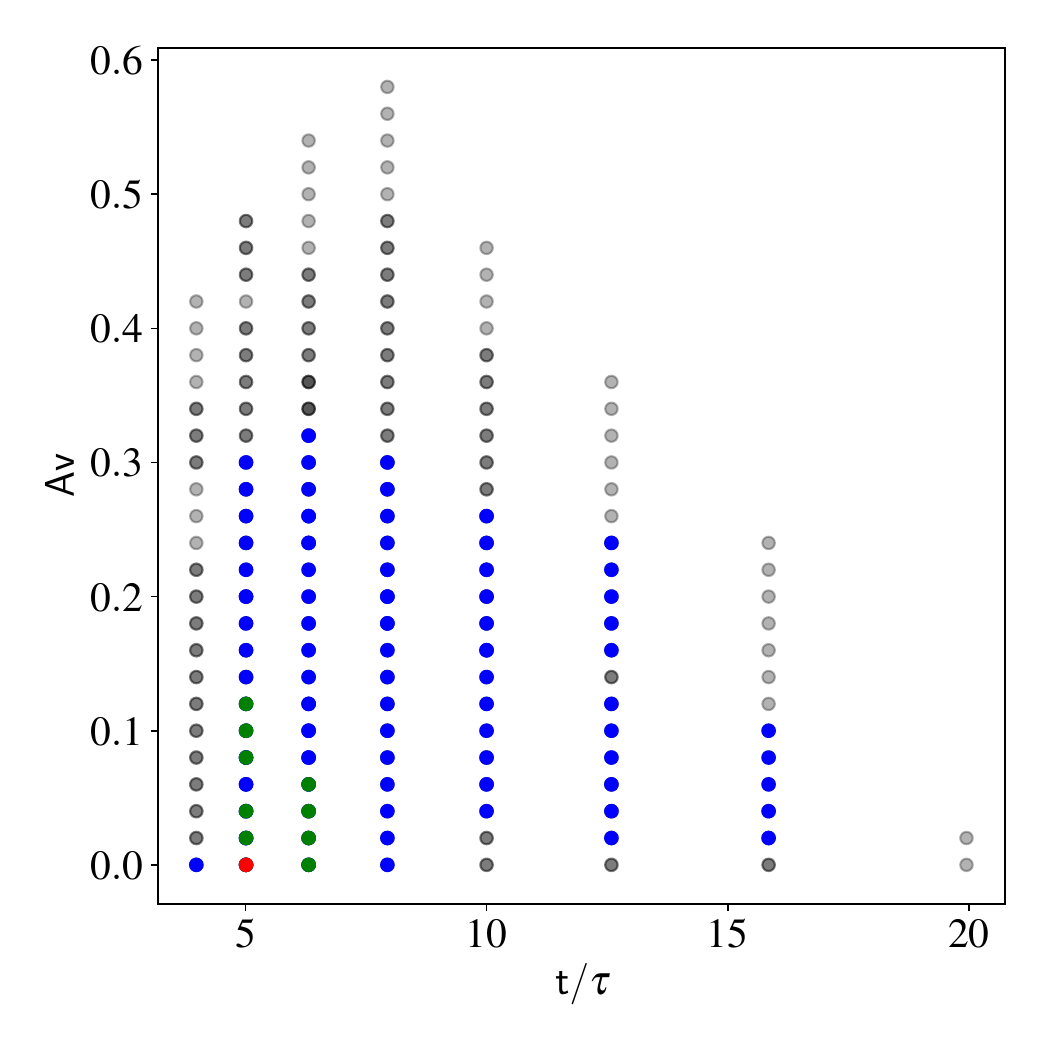}\hfill
\end{minipage}
\begin{minipage}{\textwidth}
\includegraphics[width=0.45\textwidth, clip, trim=0cm 0.0cm 0cm 0.0cm]{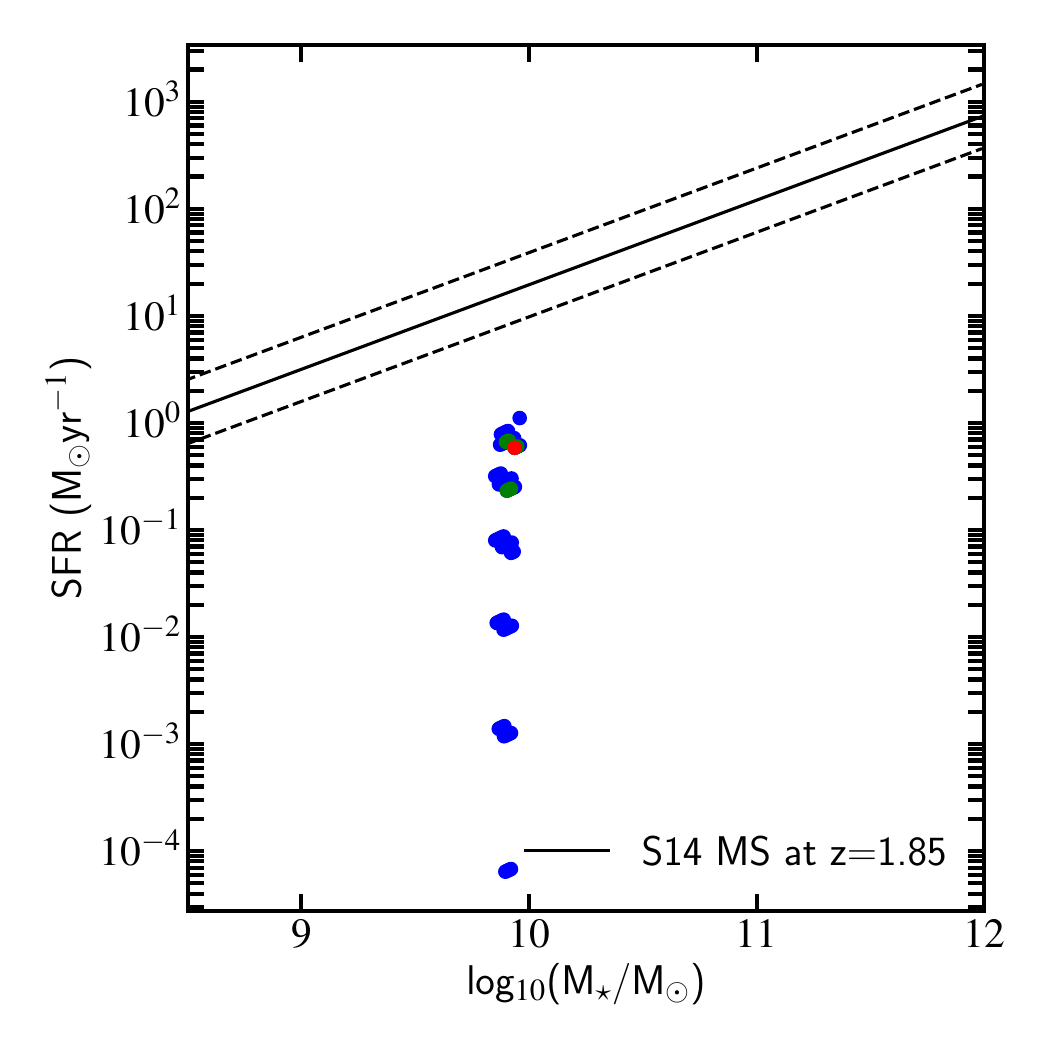}\hfill
\end{minipage}
\caption{Parameter distributions for the peak of the IHL, as labelled. Upper left: $\chi^{2}$ vs t$_{50}$ distribution for the peak of the IHL. Upper right: distribution of A$_{v}$. vs t/$\tau$. Lower left: the distribution of MS \citep{ref:M.Sargent2014} positions for the IHL peak. In all panels, grey points are within 3$\sigma$, blue points are within 2$\sigma$, green are within 1$\sigma$. The red points show the values corresponding to the minimum $\chi^{2}$.}
\label{fig:icldiagnostics2}
\end{figure*}

In Figs~\ref{fig:icldiagnostics} and \ref{fig:icldiagnostics2}, we show the outputs for FAST++ SED fitting of the peak IHL region, in terms of the distribution of output parameter values and their corresponding $\chi^{2}$ values. We also show the range of positions of the IHL peak on the MS in Fig.~\ref{fig:icldiagnostics2}, as well as the relationship between dust reddening (A$_{v}$) and $\tau$.

\section{SED comparisons}
\label{sec:sedcomp}

\begin{figure*}
\centering
\begin{minipage}{\textwidth}
\includegraphics[width=0.49\textwidth, clip, trim=0cm 0.0cm 0cm 0.0cm]{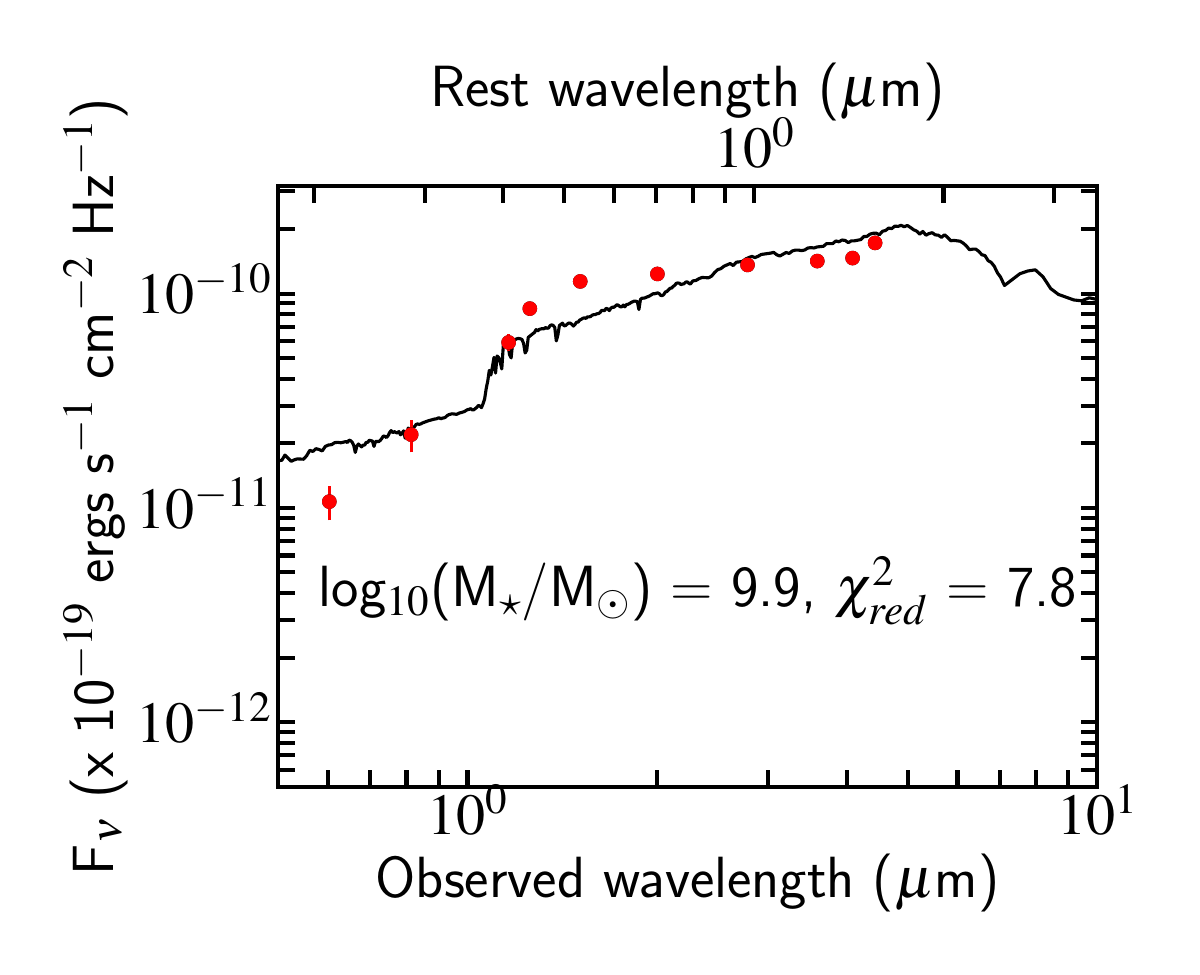}\hfill
\includegraphics[width=0.49\textwidth, clip, trim=0cm 0.0cm 0cm 0.0cm]{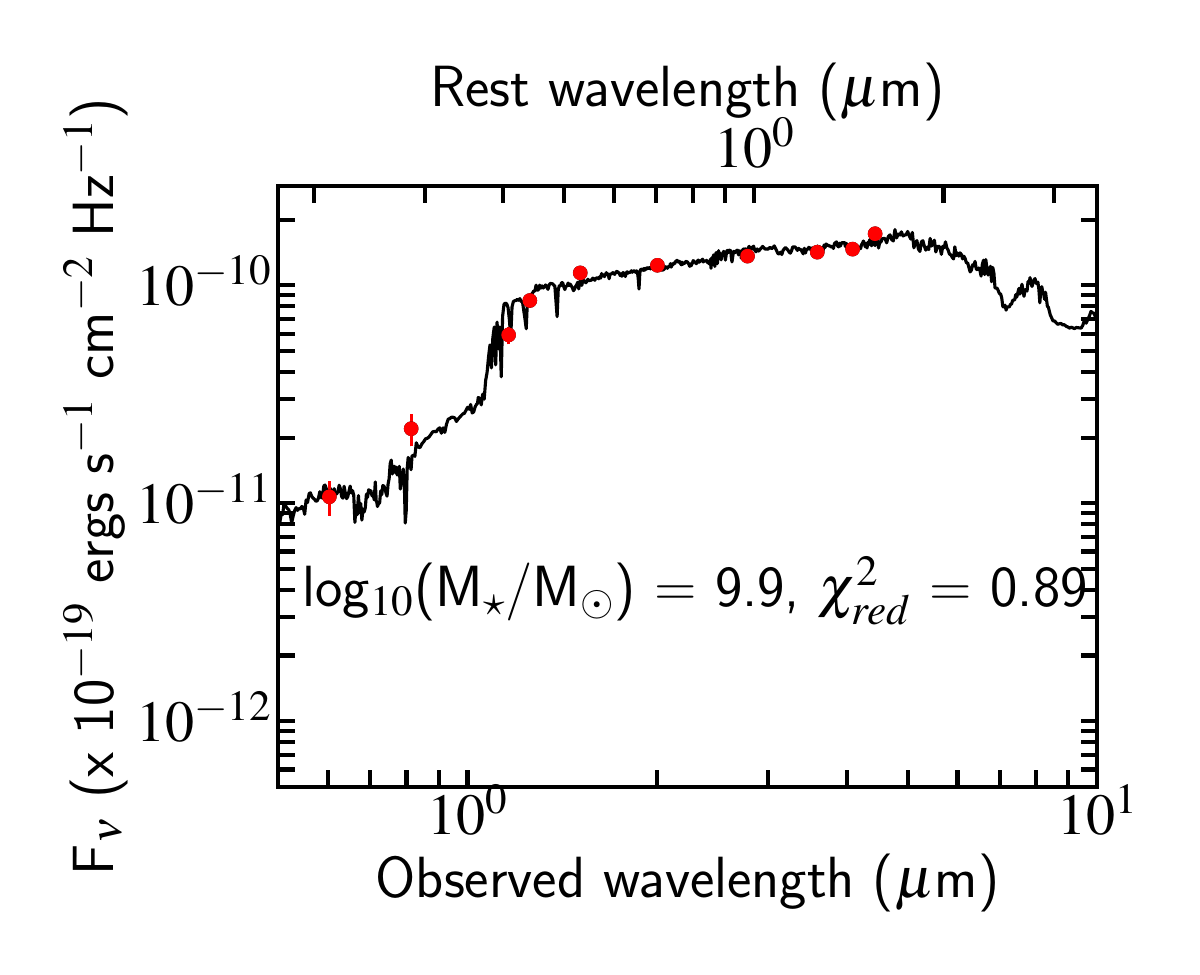}\hfill
\end{minipage}
\caption{Spectral energy distribution comparison for the IHL peak region. Left: pseudo-constant model, right: delayed-$\tau$ model.}
\label{fig:sfh_comp}
\end{figure*}

In Fig.~\ref{fig:sfh_comp}, we show the best-fitting pseudo-constant (left) and delayed-$\tau$ (right) SEDs in the IHL peak region, for comparison with the exponentially declining SFH we adopt. Whilst the pseudo-constant gives a considerably worse SED fit than the exponentially declining SFH, we see negligible difference between exponentially declining and delayed-$\tau$ models, with consistent physical parameters derived from the best-fitting SEDs.

\end{appendix}

\end{document}